\documentclass[twocolumn]{aastex63}

\usepackage{color}

\submitjournal{ApJ}

\shorttitle{Imaging Molecular Line Survey in NGC 1068}
\shortauthors{Nakajima et al.}

\begin{document}

\title{Molecular Abundance of the Circumnuclear Region Surrounding an Active Galactic Nucleus\\
in NGC 1068 based on Imaging Line Survey in the 3-mm Band with ALMA}

\correspondingauthor{Taku Nakajima}
\email{nakajima@isee.nagoya-u.ac.jp}

\author{Taku Nakajima}
\affiliation{Institute for Space-Earth Environmental Research, Nagoya University \\
Furo-cho, Chikusa-ku, Nagoya, Aichi 464-8601, Japan}

\author{Shuro Takano}
\affiliation{Department of Physics, General Studies, College of Engineering, Nihon University \\
Tamuramachi, Koriyama, Fukushima 963-8642, Japan}

\author{Tomoka Tosaki}
\affiliation{Department of Geoscience, Joetsu University of Education, \\
Yamayashiki-machi, Joetsu, Niigata 943-8512, Japan}

\author{Akio Taniguchi}
\affiliation{Division of Particle and Astrophysical Science, Graduate School of Science, Nagoya University, \\
Furo-cho, Chikusa-ku, Nagoya, Aichi 464-8601, Japan}

\author{Nanase Harada}
\affiliation{National Astronomical Observatory of Japan, \\
2-21-1, Osawa, Mitaka, Tokyo 181-8588, Japan}

\author{Toshiki Saito}
\affiliation{National Astronomical Observatory of Japan, \\
2-21-1, Osawa, Mitaka, Tokyo 181-8588, Japan}

\author{Masatoshi Imanishi}
\affiliation{National Astronomical Observatory of Japan, \\
2-21-1, Osawa, Mitaka, Tokyo 181-8588, Japan}
\affiliation{Department of Astronomy, School of Science, The Graduate University for Advanced Studies, SOKENDAI, \\
2-21-1, Osawa, Mitaka, Tokyo 181-8588, Japan}

\author{Yuri Nishimura}
\affiliation{Department of Astronomy, The University of Tokyo, \\
7-3-1 Hongo, Bunkyo-ku, Tokyo 113-0033, Japan}

\author{Takuma Izumi}
\affiliation{National Astronomical Observatory of Japan, \\
2-21-1, Osawa, Mitaka, Tokyo 181-8588, Japan}
\affiliation{Department of Physics, Graduate School of Science, Tokyo Metropolitan University, \\
1-1 Minami-Osawa, Hachioji, Tokyo 192-0397, Japan}

\author{Yoichi Tamura}
\affiliation{Division of Particle and Astrophysical Science, Graduate School of Science, Nagoya University, \\
Furo-cho, Chikusa-ku, Nagoya, Aichi 464-8601, Japan}

\author{Kotaro Kohno}
\affiliation{Institute of Astronomy, Graduate School of Science, The University of Tokyo, \\
2-21-1, Osawa, Mitaka, Tokyo 181-0015, Japan}
\affiliation{Research Center for Early Universe, Graduate School of Science, The University of Tokyo, \\
7-3-1, Hongo, Bunkyo-ku, Tokyo 113-0033, Japan}

\author{Eric Herbst}
\affiliation{Department of Chemistry, University of Virginia, \\
McCormick Road, PO Box 400319, Charlottesville, VA 22904, USA}

\begin{abstract}

We present an imaging molecular line survey in the 3-mm band (85--114 GHz) focused on one of the nearest galaxies with an active galactic nucleus (AGN), NGC 1068, based on observations taken with the Atacama Large Millimeter/submillimeter Array (ALMA). Distributions of 23 molecular transitions are obtained in the central $\sim$3 kpc region, including both the circumnuclear disk (CND) and starburst ring (SBR) with 60 and 350 pc resolution. The column densities and relative abundances of all the detected molecules are estimated under the assumption of local thermodynamic equilibrium in the CND and SBR. Then, we discuss the physical and chemical effects of the AGN on molecular abundance corresponding to the observation scale. We found that H$^{13}$CN, SiO, HCN, and H$^{13}$CO$^{+}$ are abundant in the CND relative to the SBR. In contrast, $^{13}$CO is more abundant in the SBR. Based on the calculated column density ratios of $N$(HCN)/$N$(HCO$^{+}$), $N$(HCN)/$N$(CN), and other molecular distributions, we conclude that the enhancement of HCN in the CND may be due to high-temperature environments resulting from strong shocks, which are traced by the SiO emission. Moreover, the abundance of CN in the CND is significantly lower than the expected value of the model calculations in the region affected by strong radiation. The expected strong X-ray irradiation from the AGN has a relatively lower impact on the molecular abundance in the CND than mechanical feedback.

\end{abstract}


\section{Introduction}

\begin{figure*}
\plotone{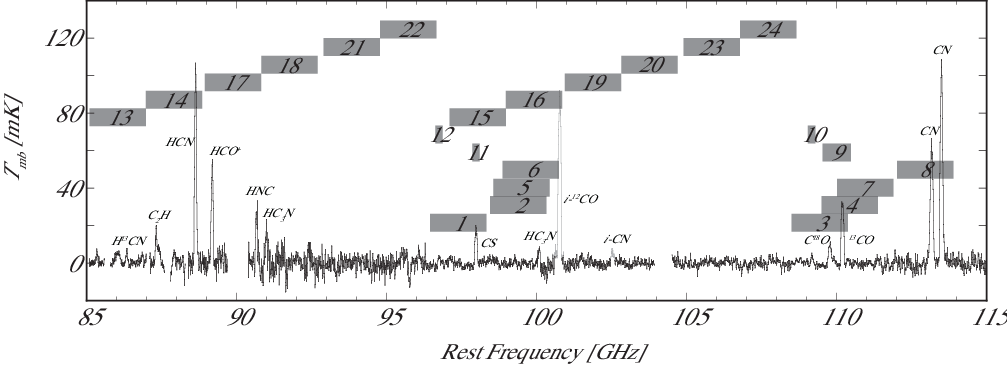}
\caption{The gray rectangles represent the frequency coverage of each spectral window, and the numbers correspond to the setup number in Table~\ref{tab:rec}. The background spectrum was obtained in NGC 1068 with the Nobeyama 45-m telescope in a previous observation \citep{tak19}.
\label{fig:spw}}
\end{figure*}

To better understand the galactic power sources associated with galaxy evolution, it is important to investigate the physical/chemical properties of a nucleus in galaxies using molecular lines, which are likely to be interstellar gas tracers free from dust extinction. Several extreme and large-scale phenomena, such as starbursts, active galactic nuclei (AGNs), and merging occur in external galaxies, unlike local objects in the Milky Way. Thus far, systematic unbiased spectral scans (i.e., molecular line surveys) have been performed for typical nuclear starbursts, AGNs, and (ultra-)luminous infrared galaxies [(U)LIRGs] using large single-dish telescopes. For example, there have been reports that focus on NGC 253, NGC 1068, NGC 7469, Arp 220, Arp 157, Mrk 231, M51, M82, M83, and NGC 3627 using the IRAM 30-m telescope \citep{mar06, dav13, ala15, wat19, qiu20}, and that focus on NGC 253, NGC 1068, IC 342, IC 10, and NGC 3627 with the NRO 45-m telescope \citep{nis16, nak18, tak19, wat19}. Details of previous line surveys of galaxies are listed in \cite{mar21}. Recently, an important and interesting issue about galaxies has been the investigation of power sources in active galaxies, focusing especially on the obscured galactic nuclei, which are the central engines of the galaxy starburst or AGN. Observations revealing the power sources may provide key information regarding the evolution of galaxies.

The chemistry-based approach, which involves the use of line surveys in galaxies, is an effective way of solving this problem. From this perspective, some observations and comparative considerations of the central regions of nearby starburst galaxies and AGNs have been carried out using single-dish telescopes. \cite{ala15} and \cite{tak19} observed the starburst galaxies and AGNs in the 3-mm band using the IRAM 30-m telescope and the NRO 45-m telescope, respectively. These studies reported that some molecules are best suited to characterize the chemistry of each type of power source. For example, large-scale shocks (e.g., CH$_{3}$OH and HNCO) and UV field (e.g., CH$_{3}$CCH and HCO) tracers dominate in starbursts, and the relative abundances of CN and HCN (and their $^{13}$C isotopologues) are enhanced in AGNs. However, the molecular abundances of the starbursts and AGNs are more similar than theoretically expected. The reason for this result in previous studies was the use of single-dish telescopes with a large beam size. For example, if the main beam (typically $\sim$kpc scale) sees a galactic center with an AGN, the emission from the starburst region surrounding the AGN is contaminated even in nearby galaxies. Therefore, observations made using a high angular resolution in the direction of the AGN are expected to reveal the pure molecular abundances without contaminations from the starburst region.

To date, several line survey observations toward nearby active galaxies with high angular resolution have been carried out using the Atacama Large Millimeter/sub-millimeter Array (ALMA). For example, the ALMA comprehensive high-resolution extragalactic molecular inventory (ALCHEMI) is an ALMA large program that investigates the starburst environment in the prototypical starburst galaxy NGC 253 \citep[e.g.,][]{mar21}. Studies of the dusty LIRG NGC 4418 \citep{cos15} and the infrared-luminous merger NGC 3256 \citep{har18} have also been reported. However, to date, there have been no fully high-resolution surveys of nearby galaxies with an AGN. Although \cite{mar15} reported some major molecular lines in the 3-mm band toward the Seyfert 1 galaxy NGC 1097, the band ranged from 85.8--89.4 GHz and 97.7--101.3 GHz, and it is not fully sampled. NGC 1068 is one of the nearest Seyfert 2 galaxies \citep[$D$=14.4 Mpc;][]{tul88}, and the central region of this galaxy consists of the circumnuclear disk (hereafter CND) and the starburst ring (hereafter SBR). This galaxy is the best target for revealing the effect of AGNs on the interstellar medium because it is easy to separate the emission from the CND, whose size is $<$300 pc \citep[$<$4$^{\prime\prime}$; e.g.,][]{sch00}, and the SBR, with a diameter of $\sim$2 kpc \citep[$\sim$30$^{\prime\prime}$; e.g.,][]{tel88}, using an interferometer. In fact, \cite{gar10}, \cite{gar14}, \cite{vit14}, \cite{gar17}, \cite{kel17}, and \cite{sco20} reported the physical conditions and chemical properties of NGC 1068 using several key molecular species observed with ALMA. 

In this study, we undertook a line survey toward NGC 1068 in the 3-mm band using ALMA as an expansion of our line survey with the NRO 45-m telescope \citep{nak18, tak19}. From the results obtained, we successfully obtained an almost complete frequency-covered spectral dataset ranging from 85 to 114 GHz by combining the four observations from cycles 0 to 2. As a result, the spatial resolution is shown to depend on the observation cycle. The largest synthesized beam is slightly smaller than 350 pc ($\sim$5$^{\prime\prime}$), which is similar to the size of the CND in NGC 1068, and the emission from the CND and SBR can be easily separated. Moreover, the highest resolution with the extended antenna configuration is approximately 60 pc ($\sim$0.$^{\prime\prime}$9) in some of the observations. The CND was resolved into a known ring structure with two bright knots, east and west, owing to the high resolution and high sensitivity of ALMA.

Here we report an imaging line survey of one of the nearest galaxies with an AGN, NGC 1068, in the 3-mm band with ALMA. To the best of our knowledge, this is the first reported high-resolution line survey that can resolve the internal structure of the CND, in a nearby galaxy that hosts an AGN. The properties of the observations and data reduction are described in Section 2. The spectra in the CND and SBR, the integrated intensity maps of each molecular line, and the estimated column densities, as well as the report of the first detected molecular lines are presented in Section 3. The results are discussed in Section 4, where we compare the fractional abundances of the CND and SBR. The conclusions of this study are listed in Section 5. In addition, we describe the calculation of rotation temperature and column density using the rotation diagram in the appendix.

\section{Observations} \label{sec:obs}

\begin{deluxetable*}{ccccccc}
\tablenum{1}
\tablecaption{Chronological summary of ALMA observations. \label{tab:obs}}
\tablewidth{0pt}
\tablehead{
\colhead{ID} & \colhead{Date} & \colhead{On source} & \colhead{ANTs} & \colhead{Configration} & \colhead{Baseline} & \colhead{$\theta_{\rm MRS}$} \\
\colhead{Session name} & \colhead{(Y/M/D)} & \colhead{(m:s)} & & & \colhead{(m)} & \colhead{(arcsec)}
}
\startdata
\multicolumn{1}{l}{2011.0.00061.S (P.I.; S. Takano)} & & & & & & \\
uid\_\_\_A002\_X369097\_X1e9.ms & 2012/01/09 & 30:38 & 16 & compact & 18.6--269.0 & 22.6 \\
uid\_\_\_A002\_X369097\_X419.ms & 2012/01/10 & 30:38 & 16 & compact & 18.6--269.0 & 22.6 \\
\multicolumn{1}{l}{2012.1.00657.S (P.I.; S. Takano)} & & & & & & \\
uid\_\_\_A002\_Xa98f9c\_X17f6.ms & 2015/09/03 & 40:08 & 36 & C34-7 & 15.1--1600 & 4.9 \\
uid\_\_\_A002\_Xa98f9c\_X239d.ms & 2015/09/03 & 40:09 & 35 & C34-7 & 15.1--1600 & 4.9 \\
\multicolumn{1}{l}{2013.1.00060.S (P.I.; T. Tosaki)} & & & & & & \\
uid\_\_\_A002\_X97aa1b\_X56e.ms & 2014/12/23 & 44:50 & 40 & C34-1/2 & 15.05--348.5 & 22.2 \\ 
uid\_\_\_A002\_X97aa1b\_X95d.ms & 2014/12/23--24 & 44:50 & 40 & C34-1/2 & 15.05--348.5 & 22.2 \\
uid\_\_\_A002\_X85c183\_X1d8c.ms & 2014/07/02 & 44:52 & 30 & C34-4 & 19.6--650.3 & 12.5 \\
uid\_\_\_A002\_X85c183\_X2341.ms & 2014/07/02 & 41:11 & 30 & C34-4& 19.6--650.3 & 12.5 \\
uid\_\_\_A002\_X87544a\_X27d0.ms & 2014/07/21 & 44:54 & 35 & C34-5 & 17.8--783.5 & 12.5 \\
uid\_\_\_A002\_Xa0b40d\_X67c5.ms & 2015/05/17 & 20:36 & 35 & C34-3 & 21.4--555.5 & 12.5 \\
\multicolumn{1}{l}{2013.1.00279.S (P.I.; T. Nakajima)} & & & & & & \\
uid\_\_\_A002\_Xa9a44e\_X13b3.ms & 2015/09/04 & 27:59 & 33 & C34-7 & 15.1--1600 & 5.1 \\
uid\_\_\_A002\_Xa9a44e\_X1865.ms & 2015/09/04 & 28:00 & 33 & C34-7 & 15.1--1600 & 5.1 \\
uid\_\_\_A002\_Xa9cdf5\_X1250c.ms & 2015/09/06 & 26:57 & 36 & C34-7 & 15.1--1600 & 5.0 \\
uid\_\_\_A002\_Xaa4256\_X2c85.ms & 2015/09/18 & 26:56 & 34 & C34-7 & 41.4--2100 & 5.0 \\
uid\_\_\_A002\_Xaa5cf7\_X54aa.ms & 2015/09/20 & 52:19 & 35 & C34-7 & 41.4--2300 & 4.0 \\
\enddata
\end{deluxetable*}

\begin{deluxetable*}{cccccccc}
\tablenum{2}
\tablecaption{Observational Parameters for Each Spectral Window\label{tab:rec}}
\tablewidth{0pt}
\tablehead{
\colhead{ID} & \multicolumn2c{Frequency range} & \multicolumn2c{$uv$ range} & \multicolumn2c{Synthesized beam} & \colhead{Rms noise} \\
\colhead{SPW No.} & \multicolumn2c{(GHz)} & \multicolumn2c{(k$\lambda$)} & \multicolumn2c{(major $\times$ minor, P.A.)} & \colhead{(mJy beam$^{-1}$)} \\
 & \colhead{start} & \colhead{end} & \colhead{min} & \colhead{max} & Briggs & natural & 
}
\startdata
2011.0.00061.S & & & & & & & \\
1 & 96.4425 & 98.3175 & 4.3 & 86.2 & 4.$\!^{\prime\prime}$2$\times$2.$\!^{\prime\prime}$4, 176$^{\circ}$ & 5.$\!^{\prime\prime}$0$\times$3.$\!^{\prime\prime}$0, 177$^{\circ}$ & 0.7 \\
2 & 98.45 & 100.325 & 4.4 & 87.9 & 4.$\!^{\prime\prime}$2$\times$2.$\!^{\prime\prime}$4, 178$^{\circ}$ & 4.$\!^{\prime\prime}$9$\times$2.$\!^{\prime\prime}$9, 177$^{\circ}$& 0.7 \\
3 & 108.5 & 110.375 & 4.8 & 96.8 & 3.$\!^{\prime\prime}$8$\times$2.$\!^{\prime\prime}$2, 178$^{\circ}$ & 4.$\!^{\prime\prime}$5$\times$2.$\!^{\prime\prime}$7, 177$^{\circ}$& 0.8 \\
4 & 109.5 & 111.375 & 4.8 & 97.6 & 3.$\!^{\prime\prime}$9$\times$2.$\!^{\prime\prime}$1, 177$^{\circ}$ & 4.$\!^{\prime\prime}$4$\times$2.$\!^{\prime\prime}$7, 177$^{\circ}$& 0.7 \\
2012.1.00657.S & & & & & & & \\
5 & 98.5625 & 100.4375 & 4.5 & 523.5 & 0.$\!^{\prime\prime}$52$\times$0.$\!^{\prime\prime}$42, 87$^{\circ}$ & 0.$\!^{\prime\prime}$59$\times$0.$\!^{\prime\prime}$57, 5$^{\circ}$& 0.4 \\
6 & 98.8625 & 100.7375 & 4.5 & 525.1 & 0.$\!^{\prime\prime}$52$\times$0.$\!^{\prime\prime}$41, 89$^{\circ}$ & 0.$\!^{\prime\prime}$59$\times$0.$\!^{\prime\prime}$58, 14$^{\circ}$& 0.4 \\
7 & 110.0125 & 111.8875 & 5.0 & 581.2 & 0.$\!^{\prime\prime}$48$\times$0.$\!^{\prime\prime}$36, 85$^{\circ}$ & 0.$\!^{\prime\prime}$53$\times$0.$\!^{\prime\prime}$51, 25$^{\circ}$& 0.4 \\
8 & 112.0125 & 113.8875 & 5.1 & 591.7 & 0.$\!^{\prime\prime}$44$\times$0.$\!^{\prime\prime}$35, 89$^{\circ}$ & 0.$\!^{\prime\prime}$52$\times$0.$\!^{\prime\prime}$50, 25$^{\circ}$& 0.4 \\
2013.1.00060.S & & & & & & & \\
9 & 109.53125 & 110.46875 & 6.7 & 233.7 & 1.$\!^{\prime\prime}$04$\times$0.$\!^{\prime\prime}$93, 149$^{\circ}$ & 1.$\!^{\prime\prime}$75$\times$1.$\!^{\prime\prime}$43, 68$^{\circ}$& 0.1 \\
10 & 109.05643 & 109.29083 & 6.6 & 231.2 & 1.$\!^{\prime\prime}$09$\times$0.$\!^{\prime\prime}$95, 158$^{\circ}$ & 1.$\!^{\prime\prime}$75$\times$1.$\!^{\prime\prime}$44, 70$^{\circ}$& 0.2 \\
11 & 97.863753 & 98.098153 & 6.0 & 207.5 & 1.$\!^{\prime\prime}$22$\times$1.$\!^{\prime\prime}$06, 158$^{\circ}$ & 1.$\!^{\prime\prime}$97$\times$1.$\!^{\prime\prime}$55, 65$^{\circ}$& 0.2 \\
12 & 96.624175 & 96.858575 & 5.1 & 204.9 & 1.$\!^{\prime\prime}$23$\times$1.$\!^{\prime\prime}$08, 157$^{\circ}$ & 1.$\!^{\prime\prime}$93$\times$1.$\!^{\prime\prime}$60, 70$^{\circ}$& 0.1 \\
2013.1.00279.S (Set-1) & & & & & & & \\
13 & 85.1 & 86.975 & 4.3 & 411.4 & 0.$\!^{\prime\prime}$65$\times$0.$\!^{\prime\prime}$53, 84$^{\circ}$ & 0.$\!^{\prime\prime}$87$\times$0.$\!^{\prime\prime}$68, 57$^{\circ}$& 0.3 \\
14 & 86.975 & 88.85 & 4.4 & 420.3 & 0.$\!^{\prime\prime}$65$\times$0.$\!^{\prime\prime}$50, 58$^{\circ}$ & 0.$\!^{\prime\prime}$87$\times$0.$\!^{\prime\prime}$66, 57$^{\circ}$ & 0.4 \\
15& 97.1 & 98.975 & 4.9 & 464.2 & 0.$\!^{\prime\prime}$55$\times$0.$\!^{\prime\prime}$47, 84$^{\circ}$ & 0.$\!^{\prime\prime}$79$\times$0.$\!^{\prime\prime}$59, 59$^{\circ}$& 0.4 \\
16 & 99.975 & 100.85 & 4.9 & 473.1 & 0.$\!^{\prime\prime}$54$\times$0.$\!^{\prime\prime}$47, 85$^{\circ}$ & 0.$\!^{\prime\prime}$77$\times$0.$\!^{\prime\prime}$58, 59$^{\circ}$& 0.4 \\
2013.1.00279.S (Set-2) & & & & & & & \\
17 & 88.95 & 90.825 & 3.9 & 539.6 & 0.$\!^{\prime\prime}$53$\times$0.$\!^{\prime\prime}$39, 86$^{\circ}$ & 0.$\!^{\prime\prime}$65$\times$0.$\!^{\prime\prime}$55, 126$^{\circ}$& 0.3 \\
18 & 90.825 & 92.7 & 4.0 & 550.9 & 0.$\!^{\prime\prime}$54$\times$0.$\!^{\prime\prime}$39, 85$^{\circ}$ & 0.$\!^{\prime\prime}$65$\times$0.$\!^{\prime\prime}$54, 124$^{\circ}$ & 0.3 \\
19 & 100.95 & 102.825 & 4.4 & 605.9 & 0.$\!^{\prime\prime}$50$\times$0.$\!^{\prime\prime}$35, 84$^{\circ}$ & 0.$\!^{\prime\prime}$57$\times$0.$\!^{\prime\prime}$48, 117$^{\circ}$ & 0.4 \\
20 & 102.825 & 104.7 & 4.4 & 617.1 & 0.$\!^{\prime\prime}$45$\times$0.$\!^{\prime\prime}$32, 81$^{\circ}$ & 0.$\!^{\prime\prime}$56$\times$0.$\!^{\prime\prime}$47, 118$^{\circ}$ & 0.5 \\
2013.1.00279.S (Set-3) & & & & & & & \\
21 & 92.9 & 94.775 & 11.4 & 557.6 & 0.$\!^{\prime\prime}$46$\times$0.$\!^{\prime\prime}$34, 77$^{\circ}$ & 0.$\!^{\prime\prime}$70$\times$0.$\!^{\prime\prime}$45, 57$^{\circ}$ & 0.4 \\
22 & 94.775 & 96.65 & 11.6 & 568.7 & 0.$\!^{\prime\prime}$48$\times$0.$\!^{\prime\prime}$34, 76$^{\circ}$ & 0.$\!^{\prime\prime}$68$\times$0.$\!^{\prime\prime}$44, 57$^{\circ}$ & 0.4 \\
23 & 104.9 & 106.775 & 12.7 & 623.2 & 0.$\!^{\prime\prime}$41$\times$0.$\!^{\prime\prime}$31, 74$^{\circ}$ & 0.$\!^{\prime\prime}$63$\times$0.$\!^{\prime\prime}$40, 57$^{\circ}$ & 0.5 \\
24 & 106.75 & 108.65 & 12.9 & 634.3 & 0.$\!^{\prime\prime}$40$\times$0.$\!^{\prime\prime}$31, 73$^{\circ}$ & 0.$\!^{\prime\prime}$61$\times$0.$\!^{\prime\prime}$39, 57$^{\circ}$ & 0.5 \\
\enddata
\end{deluxetable*}

We carried out line survey observations in the 3-mm band toward NGC 1068 with ALMA. Line survey data were obtained in combination with four observation programs from cycle-0 in 2012 to cycle-2 in 2015, with IDs of 2011.0.00061.S (P.I.; S. Takano), 2012.1.00657.S (P.I.; S. Takano), 2013.1.00060.S (P.I.; T. Tosaki), and 2013.1.00279.S (P.I.; T. Nakajima). The observational parameters are summarized in Table~\ref{tab:obs}. The number of 12-m antennas and the layout were 16 in a compact configuration in cycle-0 as the early science phase, while 30--40 antennas with longer baselines were used in cycles-1 and 2. The results of the first observations (2011.0.00061.S) were published by \citet{tak14} and \citet{nak15}. The second observation period (2012.1.00657.S) was proposed and accepted in cycle-1; in fact, the observations were carried out in the period of cycle-2, which was the same as the fourth observation (2013.1.00279.S). Therefore, the observational parameters were quite similar between these two observations; in particular, the antenna layout was an extended configuration, and the maximum baseline was more than 1600 m. The maximum recoverable scales ($\theta_{\rm MRS}$) of these two observations were smaller than those of the other observations, and these values were approximately five arcsec. The convolved angular resolutions of this line survey (0.$\!^{ \prime\prime }$9 and 5.$\!^{ \prime\prime }$0) were only smaller than this scale, except for the highest frequency range of the fourth observation ($\theta_{\rm MRS}$ = 4 arcsec). Note that only the spectrum of N$_{2}$H$^{+}$ is located in this highest frequency range and that the flux density can be slightly underestimated. The third observation (2013.1.00060.S) was used for both the compact and extended antenna configurations, and the results based on these observations were reported by \citet{tos17}.

The observation setup is presented in Table~\ref{tab:rec}. Because the 2013.1.00279.S observations have three science goals, they are divided into three sets from Set-1 to Set-3 in this table. In total, 24 spectral windows (SPWs) cover the 3-mm region from 85.1 to 113.9 GHz. The frequency range of each SPW is illustrated on the spectrum obtained in NGC 1068 with the NRO 45-m single-dish telescope \citep{tak19} in Figure~\ref{fig:spw}. All observations were performed using a band-3 receiver (84--116 GHz); the two SPWs were placed in the lower sideband, and the other two were placed in the upper sideband for the dual-polarization sideband-separating receiver. Each SPW had a range of 1875 MHz with 3840 channels, resulting in frequency and velocity resolutions of 488 kHz and 2.6--3.9 km s$^{-1}$, respectively, except for SPW numbers Nos.9--12. The velocity resolution was 1.3--1.5 km s$^{-1}$ for Nos.9--12 observations, but the frequency bandwidth was narrow relative to other observations. Details of the parameters are presented in \citet{tos17}. In this study, the results are presented with a velocity resolution of $\sim$20 km s$^{-1}$ to improve the signal-to-noise ratio (S/N). The synthesized beam with Briggs weighting was $\sim$4$^{\prime\prime}$ $\times$ 2$^{\prime\prime}$ ($\sim$280 pc $\times$ 140 pc at the adopted distance of NGC 1068) for Nos.1--4, $\sim$1$^{\prime\prime}$ ($\sim$70 pc) for Nos.9--12, and $\sim$0.5$^{\prime\prime}$ ($\sim$35 pc) for Nos.5--8 and Nos.13--24. The achieved noise level (RMS) was approximately 0.1--0.5 mJy beam$^{-1}$ at a velocity resolution of 20 km s$^{-1}$.

The phase reference center was set to $\alpha_{J2000}$ = 2$^{h}$42$^{m}$40.$\!^{s}$798 and $\delta_{J2000}$ = -00$^{\circ}$00$^{\prime}$47$^{\prime\prime}$938 \citep{sch00}, which corresponds to the radio position of the active nucleus \citep{mux96} for 2011.0.00061.S and 2012.1.00657.S. In 2013.1.00060.S and 2013.1.00279.S, the position was set to $\alpha_{J2000}$ = 2$^{h}$42$^{m}$40.$\!^{s}$709 and $\delta_{J2000}$ = -00$^{\circ}$00$^{\prime}$47$^{\prime\prime}$945 \citep{gal04}. The systemic velocity employed was 1150 km s$^{-1}$ for all of the observations. The 12-m antenna primary beams in this receiver range from HPBW = 53$^{\prime\prime}$ to 74$^{\prime\prime}$ for 84--116 GHz.  Therefore, the circumnuclear disk (CND) with a $\lesssim$280-pc ($\lesssim$4$^{\prime\prime}$) diameter as well as surrounding starburst ring (SBR) with a  $\sim$2-kpc ($\sim$30$^{\prime\prime}$) diameter in NGC 1068 were covered and simultaneously imaged by performing single pointing observations. 

The calibration and imaging of the delivered data provided by the ALMA Regional Center were carried out using Common Astronomy Software Applications (CASA)\footnote{https://casa.nrao.edu} \citep{mcm07}. We used the CASA task {\tt\string uvcontsub} to subtract a continuum emission as a linear baseline based on the level of emission-free channels from the calibrated visibilities and {\tt\string tclean} for image reconstruction and deconvolution with natural weighting. Although this weighting function produced images with the poorest angular resolution, we expected to obtain the highest S/N spectra for the molecular line survey. The resulting spatial resolution is listed in the 7th column of Table~\ref{tab:rec}.

\section{Results}
\subsection{New molecular line detections}

\begin{figure*}
\plotone{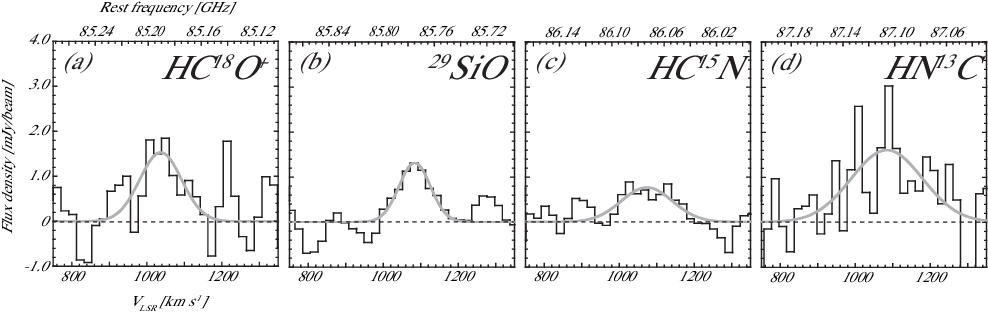}
\caption{The spectra and Gaussian fitting curves of the first detection or the first observation with the interferometer in NGC 1068 include (a) HC$^{18}$O$^{+}$, (b) $^{29}$SiO, (c) HC$^{15}$N, and (d) HN$^{13}$C. The vertical axis is the flux density per convolved beam. The spatial resolution is 5.$\!^{\prime\prime}$0 ($\sim$350 pc) for (a) and (d) centered at the AGN position, and 0.$\!^{\prime\prime}$9 ($\sim$60 pc) for (b) and (c) centered at the E-knot position (coordinates are explained in the text of section 3.2). The horizontal axes represent the velocity ($V_{\rm LSR}$; lower) and rest frequency (upper). 
\label{fig:spectra}}
\end{figure*}

Figure ~\ref{fig:spectra} shows spectra of the first detection or of new detections with the interferometer in NGC 1068. The molecules detected include (a) HC$^{18}$O$^{+}$ ($J$ = 1--0), (b) $^{29}$SiO ($J$ = 2--1), (c) HC$^{15}$N ($J$ = 1--0), and (d) HN$^{13}$C ($J$ = 1--0). Profiles (a) and (d) were obtained within the circular region with a 5$^{\prime\prime}$ diameter at the AGN position, which covers approximately the entire structure of the CND. The integrated fluxes of the detected lines are approximately 3$\sigma$ of the noise level. Profiles (b) and (c) were obtained at the E-knot position, with a resolution of 0.$\!^{\prime\prime}$9, and the detection levels are more than 4$\sigma$ for these lines.

\cite{wan14} and \cite{qiu18} have already reported the detection of HC$^{18}$O$^{+}$ toward NGC 1068 with the IRAM 30-m single-dish telescope. The calculated peak velocity with Gaussian fitting obtained using ALMA is 1037$\pm$27 km s$^{-1}$ (Figure ~\ref{fig:spectra}(a)), and this value is consistent with that in previous studies. In addition, the line shape, which has several velocity components, is almost identical to that of the IRAM 30-m.

\cite{qiu18} reported the detection of H42$\alpha$ at 85.6950 GHz; however, it was not detected around that frequency in this observation. We found line features in the range of 85.75--85.80 GHz (Figure ~\ref{fig:spectra}(b)), which are close to the frequency of this recombination line. This range is in close agreement with the frequency of $^{29}$SiO \citep[$J$ = 2--1, v=0; 85.759144 GHz;][]{lov04}, and the detection of this isotopologue in NGC 1068 is the first time such detection.

Extragalactic HC$^{15}$N has been reported in the Large Magellanic Cloud (LMC) and composite AGN-starburst galaxy NGC 4945 \citep{chi99}. However, in NGC 1068, previous line survey observations in the 3-mm band with single-dish telescopes have not yet detected this molecular line \citep{ala15, qiu18, tak19}. The only upper limit reported is $<$115 mK km s$^{-1}$ (2$\sigma$) by \cite{wan14}. In this observation, HC$^{15}$N was detected in the peak positions of both the E- and W-knots, which were detected at 3.5--4$\sigma$. The integrated line ratios of HC$^{14}$N/HC$^{15}$N in LMC and NGC 4945 were reported by \cite{chi99} to be 49--69 and 46, respectively. In this work, this ratio at the E-knot position in NGC 1068 is calculated as 64.8$\pm$15.0, which is close to that in the LMC.

HN$^{13}$C in NGC 1068 has been reported by \cite{wan14} and \cite{qiu18} using the IRAM 30-m telescope. In this study, the integrated flux ratio of HN$^{12}$C/HN$^{13}$C was measured to be 26.4$\pm$8.5, which ranges between 23.3 \citep{ala13} and 38.1 \citep{wan14}. This consistency indicates that HN$^{13}$C is mainly located in the CND with a minor contribution from the SBR.

\subsection{Molecular gas distributions}

\begin{figure*}
\plotone{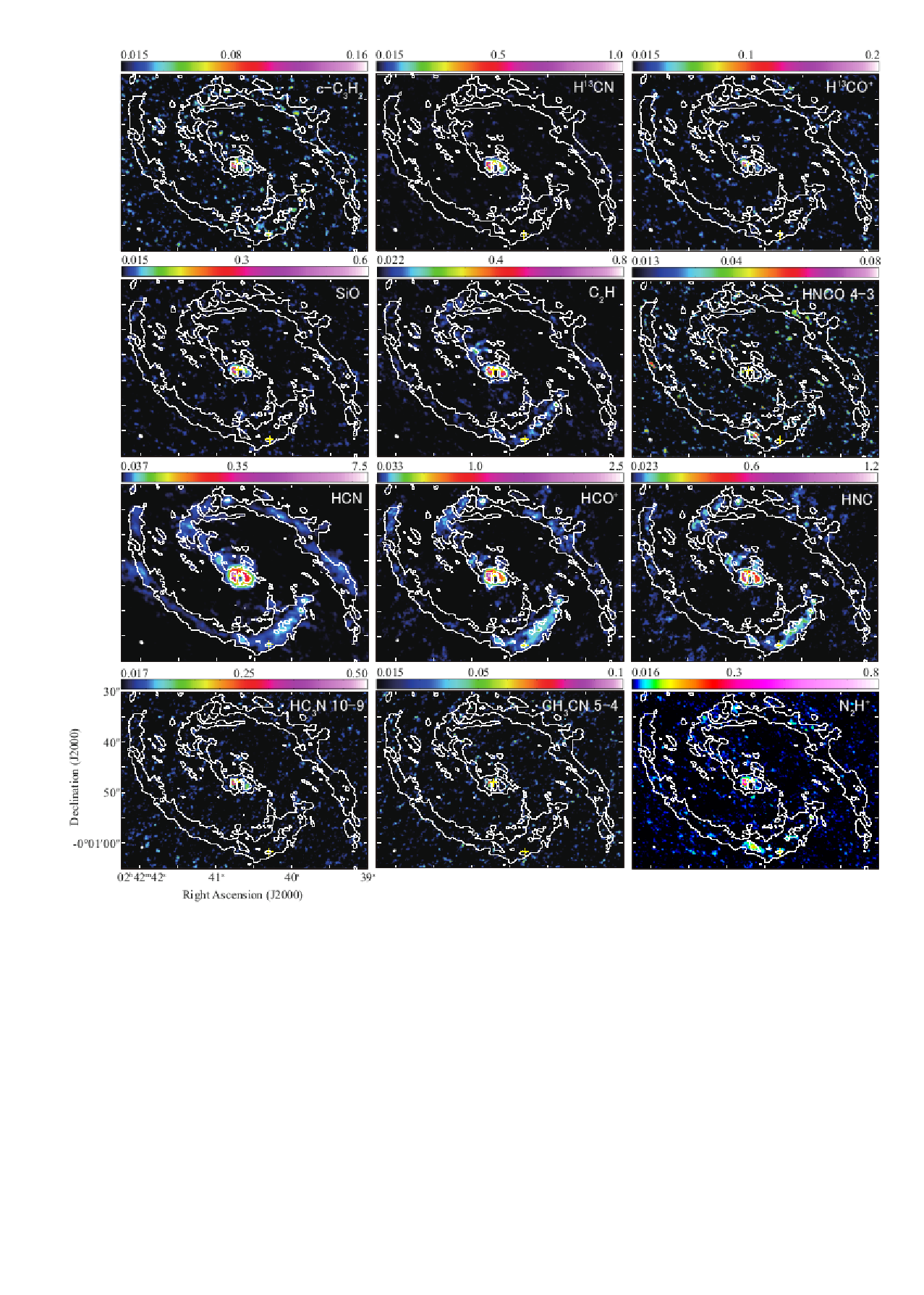}
\caption{The images of integrated flux (moment-0) for 23 molecular lines, which are clearly detected both in the CND and the SBR in NGC 1068. The unit of the color scale is Jy beam$^{-1}$ km s$^{-1}$, and the minimum color levels are set to 1$\sigma$ noise level for each molecular line. The contour maps are 3$\sigma$ and 9$\sigma$ of $^{12}$CO ($J$ = 1--0); the integrated flux indicates a typical gas distribution. The positions of the AGN \citep{gal04} and the southwestern region in the SBR \citep{tak14} are indicated using yellow cross marks.
\label{fig:mom0a}}
\end{figure*}

\begin{figure*}
\plotone{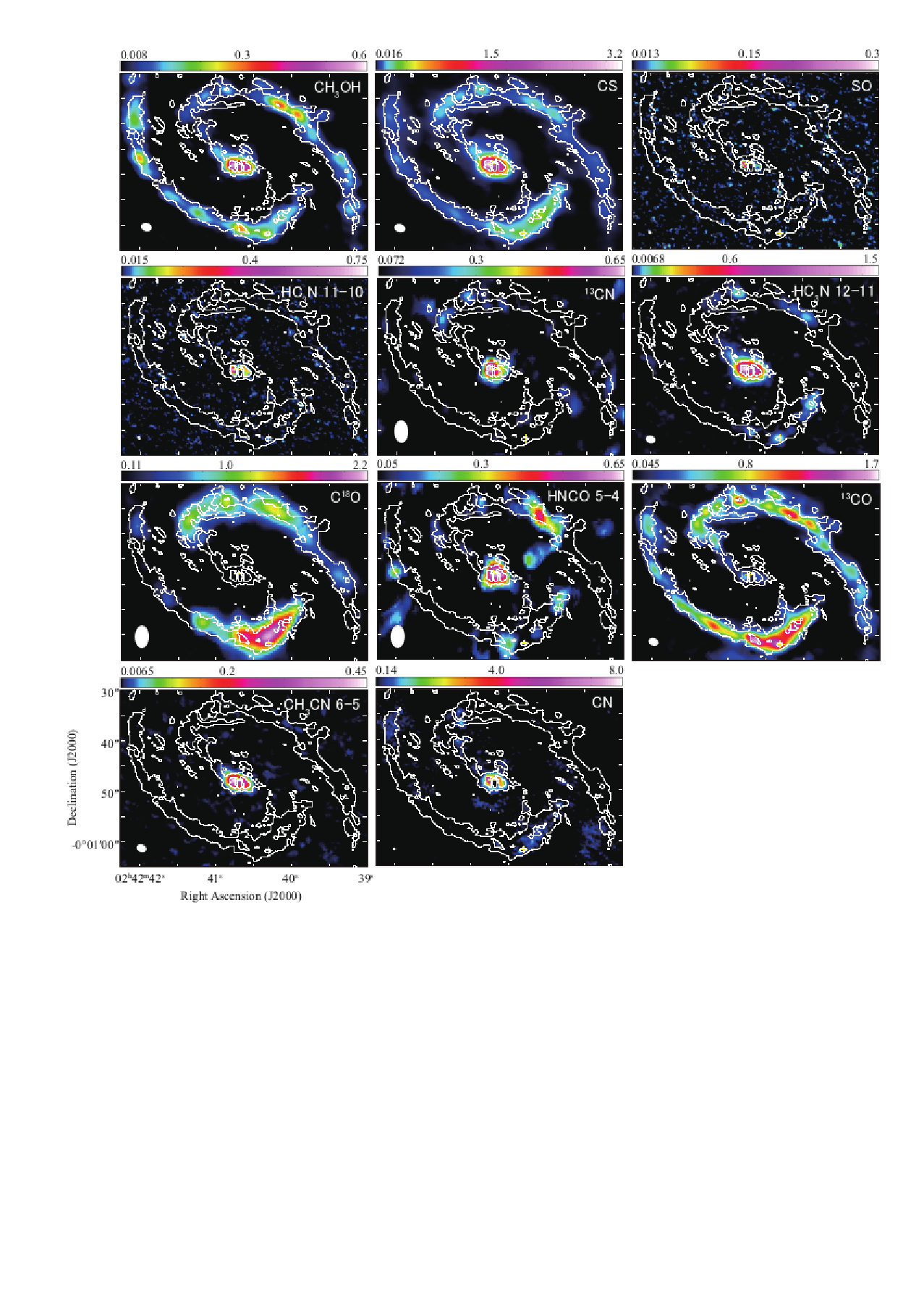}
\caption{(continued).
\label{fig:mom0b}}
\end{figure*}

\begin{figure*}
\plotone{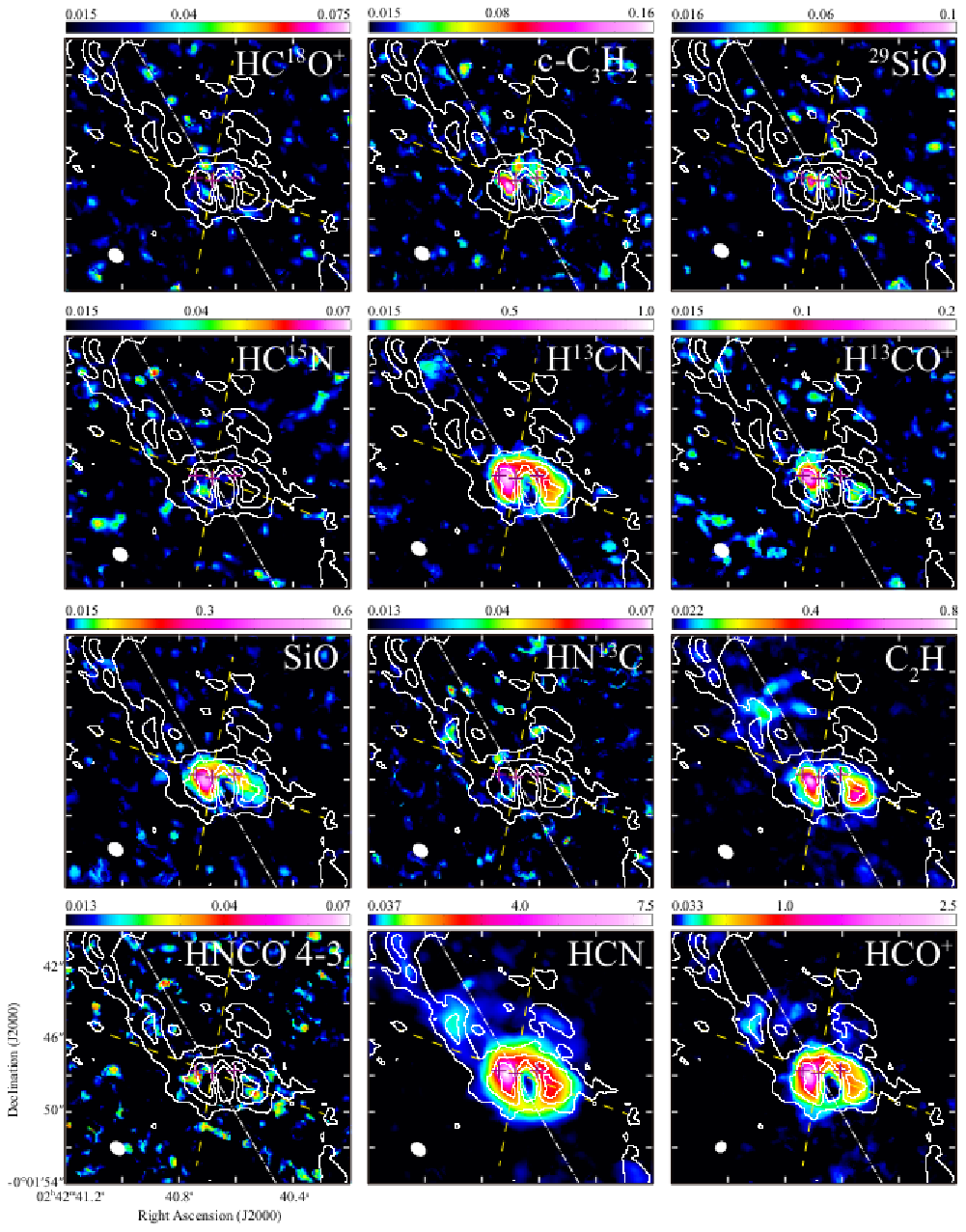}
\caption{The enlarged integrated flux (moment-0) view of the CND for 23 molecular lines, which are obtained with high spectral resolution (synthesized beam$<$1$^{\prime\prime}$) and include the newly detected lines (see Section 3.1). The unit of the color scale is Jy beam$^{-1}$ km s$^{-1}$, and the minimum color levels are set to a 1$\sigma$ noise level for each molecular line. The contour maps are 3$\sigma$, 6$\sigma$, and 12$\sigma$ of $^{12}$CO ($J$ = 1--0); the integrated flux indicates a typical gas distribution. Three magenta cross marks represent the positions of the E-knot, AGN, and W-knot from the left side (the coordinates are explained in the text of Section 3.2). The dual cone structure with a 30$^{\circ}$ position angle (dash-dot-dash line) and 40$^{\circ}$ outer opening angle (dashed-line) from the location of the supermassive black hole is represented in all images \citep{das06}.
\label{fig:mom0c}}
\end{figure*}

\begin{figure*}
\plotone{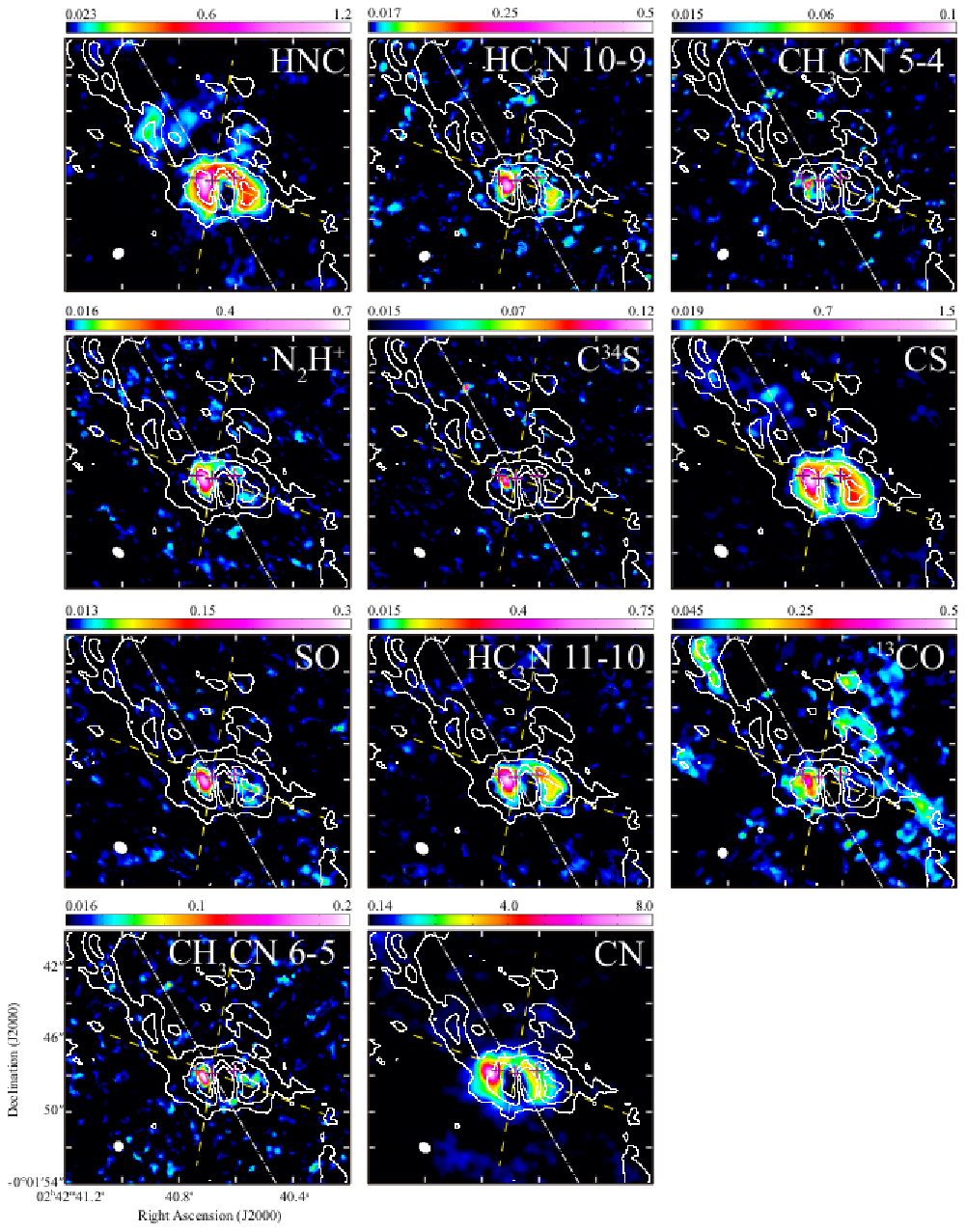}
\caption{(continued).
\label{fig:mom0d}}
\end{figure*}

\begin{figure*}
\plotone{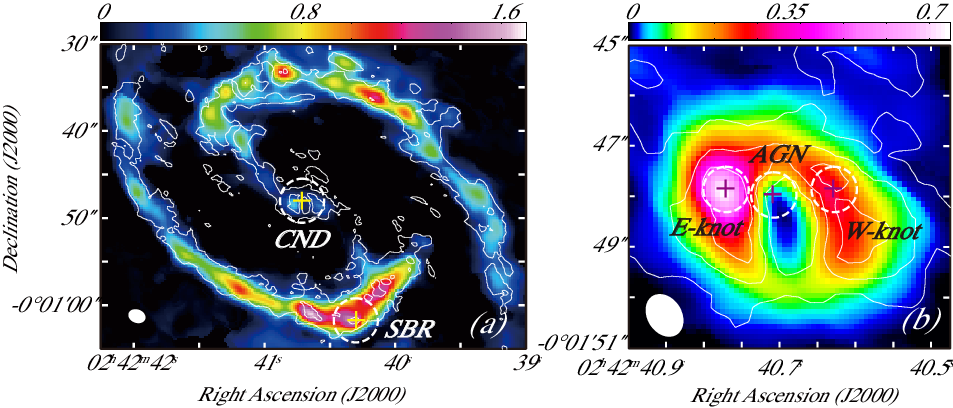}
\caption{The extracted positions of spectral data are indicated by crosses on (a) the $^{13}$CO moment-0 map for the CND and SBR, and on (b) the HCN moment-0 map for the AGN, E-, and W-knots. These coordinates are explained in the text of Section 3.2. The contour map shows a $^{12}$CO distribution (see the captions of Figures~\ref{fig:mom0a} and ~\ref{fig:mom0c}). The dashed-line circles represent the convolved beam sizes of 5.$\!^{\prime\prime}$0 ($\sim$350 pc) in (a) and 0.$\!^{\prime\prime}$9 ($\sim$60 pc) in (b). These beams do not overlap and the samples are spatially independent.
\label{fig:beam}}
\end{figure*}

Based on the molecular lines detected in our previous line survey observation with a single-dish telescope \citep{tak19}, we obtained the spatial distributions of the major molecular species with ALMA. In this ALMA observation, which ranged from 84--114 GHz, we successfully obtained 21 molecular line distributions for all detected molecular lines using the NRO 45-m telescope. Moreover, non-detected lines with the NRO 45-m telescope, H$^{13}$CO$^{+}$ ($J$ = 1--0) and CH$_{3}$CN ($J_{K}$ = 5$_{K}$--4$_{K}$), were clearly detected and imaged using ALMA. Figures~\ref{fig:mom0a} and ~\ref{fig:mom0b} show the integrated flux (moment-0) as color maps of 23 molecular lines (15 molecular species and four isotopologues) covering the entire central structure, which consists of the CND and SBR. The positions of the AGN \citep{gal04} and southwestern region in the SBR \citep{tak14} are indicated using cross marks on all images. The minimum color level for each molecular line was set to a 1$\sigma$ noise level. These levels were estimated from the averaged RMS noise in the region of the emission-free blank sky as typical noise over the map. The overlaid contour map in each image represents the integrated flux of $^{12}$CO ($J$ = 1--0) obtained during observations 2018.1.01684.S (P.I. T. Tosaki) which exhibited a typical gas distribution. These contours represent the 3$\sigma$ and 9$\sigma$ noise levels. All $uv$ range data are used in the imaging process, and the range is different on an observational basis (see Table~\ref{tab:rec}). Primary beam correction was not applied to any of the integrated flux images (Figures~\ref{fig:mom0a}--\ref{fig:mom0d}).

A method was applied to create integrated flux maps with higher S/N ratios compared to the standard process using the {\tt\string immoments} task in CASA. The central channel (i.e., the peak velocity) and the number of channels for integration (i.e., the velocity width) of the spectrum are the same for each pixel in the simplest method. However, in the new method, the values are changed on a pixel-by-pixel basis depending on the peak velocity (moment-1) and velocity dispersion ($\sigma$$V$; moment-2) of $^{12}$CO ($J$ = 1--0) as the template. The detailed procedure of this method will be presented in a separate paper (A. Taniguchi et al. in prep.). The velocity width for integration was calculated using the velocity dispersion of the template spectrum multiplied by an arbitrary factor, and we applied 3$\sigma$$V$ in this work. The peak velocity and velocity width applied to the integrated flux map images are different for each pixel, but these values are common for all molecular lines. For example, the typical peak velocity at the CND is 1100--1150 km s$^{-1}$. Those in the eastward and westward SBR regions from the CND are approximately 1000--1050 km s$^{-1}$ and 1200--1250 km s$^{-1}$, respectively. The typical velocity dispersion in the CND and SBR regions was approximately 80 km s$^{-1}$ and 50 km s$^{-1}$, respectively. 

Although the distributions of $^{13}$CN, C$^{18}$O, and HNCO ($J_{Ka, Kc}$ = 5$_{0,5}$--4$_{0,4}$) were reported in \cite{tak14}, and CH$_{3}$OH, CS, and $^{13}$CO were already reported in \cite{tos17} using the same data, these images are presented again in this paper for completeness. The S/N of these images was drastically improved owing to the new method. In addition, images of SO, HC$_{3}$N ($J$ = 11--10; 12--11), and CH$_{3}$CN ($J_{K}$ = 6$_{K}$--5$_{K}$), which are taken in the cycle-0 period, have been reported by \cite{tak14}. However, these images are presented in this paper again using the newly obtained data in cycle-2 because the spatial resolutions are approximately ten times higher than those of the previous works.

To date, images of the molecular distribution using interferometers covering both the CND and SBR in NGC 1068 have been previously reported. However, these were obtained only for relatively strong emission lines such as those of CO, HCN, HCO$^{+}$, and CS. For $^{12}$CO see \cite{pla91}, \cite{kan92}, \cite{hel95}, \cite{sch00}, \cite{tsa12}; for $^{13}$CO see \cite{hel95}, \cite{pap96}, \cite{tac97}; for C$^{18}$O see \cite{pap96}. Other references include \cite{tac94}, \cite{hel95}, \cite{koh08} for HCN, \cite{koh08} for HCO$^{+}$, and \cite{tac97}, \cite{sco20} for CS. Observations of other molecules toward this galaxy are quite limited even in the 3-mm band. With the exception of papers reported by our group \citep{tak14, tos17, sai22a, sai22b}, SiO \citep{gar10, hua22}, C$_{2}$H \citep{gar17}, CS \citep{sco20}, and HCN, HCO$^{+}$ \citep{san22} with ALMA are the only reported studies. Therefore, many of the molecular distribution images obtained in the central region of NGC 1068 are being presented for the first time.  

In Figures~\ref{fig:mom0a} and ~\ref{fig:mom0b}, the molecules showing a variety of spatial distributions are likely due to the difference in physical/chemical properties between the CND and SBR. The strong emissions of H$^{13}$CN, H$^{13}$CO$^{+}$, SiO, HC$_{3}$N, CH$_{3}$CN, SO, and $^{13}$CN are concentrated in the CND, whereas weak emissions were observed in the SBR. The enhancement of these molecules in the CND may be due to the effect of the AGN. In contrast, C$^{18}$O and $^{13}$CO were mainly distributed in the SBR, and were significantly weak in the CND. This feature is well known from previous studies, and we have already discussed it in \cite{tak14}. Other molecules, cyclic-C$_{3}$H$_{2}$ (hereafter $c$-C$_{3}$H$_{2}$), C$_{2}$H, HNCO, HCN, HCO$^{+}$, HNC, N$_{2}$H$^{+}$, CH$_{3}$OH, CS, and CN, are distributed in both the CND and SBR. We find that the emissions of $c$-C$_{3}$H$_{2}$, HNCO ($J_{Ka, Kc}$ = 4$_{0,4}$--3$_{0,3}$), and N$_{2}$H$^{+}$ in the SBR show particularly clumpy structures without diffuse components at the peak positions of $^{12}$CO. However, this may be due to an insufficient S/N or missing flux because these emission lines are very weak, especially in the SBR relative to other molecules in this category. In addition, CH$_{3}$OH ($J_{K}$ = 2$_{K}$--1$_{K}$) and HNCO ($J_{Ka, Kc}$ = 5$_{0,5}$--4$_{0,4}$) exhibit a characteristic distribution in the SBR. These peak positions are clearly different from those of $^{12}$CO. These molecules may reflect dynamic effects, such as cloud-cloud collisions and/or galaxy dynamics, instead of interstellar mass distribution, because they are known as shock tracers. The distribution of CH$_{3}$OH was also analyzed and discussed in \cite{tos17}.

Figures ~\ref{fig:mom0c} and ~\ref{fig:mom0d} show enlarged views of the CND. We obtained a clear distribution of 23 molecular lines with a peak S/N$>$5 and a $<$1.0 arcsec resolution. The image size is approximately 15$^{\prime\prime}$ square ($\sim$1 kpc at a distance of 14.4 Mpc), which is almost the same size as the NRO 45-m telescope main beam around the center of the CND. The three crosses represent the positions of the E-knot, AGN, and W-knot, which were defined by \cite{gal04} for the AGN and \cite{vit14} for the others, from the left side of each image. In addition, the configuration of the biconical outflow model \citep{das06} is indicated in all images. They proposed a biconical structure with a 30$^{\circ}$ position angle and 40$^{\circ}$ outer opening angle from the location of the supermassive black hole based on observations of the narrow-line region in NGC 1068 obtained with the Hubble Space Telescope (HST) and the Multi-Element Radio Linked Interferometer Network (MERLIN) radio maps. We expect to find dynamical and chemical effects owing to the interaction between the AGN jet-driven outflow and the galactic disk, as proposed by \cite{gar14} and \cite{sai22a}.

Some molecular lines, H$^{13}$CN, SiO, C$_{2}$H, HCN, HCO$^{+}$, HNC, CS, and CN, show a clear ring structure in the CND, similar to the $^{12}$CO distribution (i.e., the contours). Other molecules were mainly detected in the vicinity of the E- and W-knots. C$_{2}$H, HC$_{3}$N, and CS show comparable peaks in both knots, but the following molecules are stronger in the E-knot with moderate detections in the W-knot: c-C$_{3}$H$_{2}$, $^{29}$SiO, H$^{13}$CO$^{+}$, CH$_{3}$CN, N$_{2}$H$^{+}$, C$^{34}$S, SO, and $^{13}$CO. There is no stronger molecular line in the W-knot as in the exceptional case of a neutral atomic carbon line \citep[see][]{sai22a}. The peak intensity position in the W-knot varies with the molecule. In particular, H$^{13}$CO$^{+}$, HN$^{13}$C, C$_{2}$H, HNCO, HC$_{3}$N ($J$ = 10--9), and SO peaks are approximately 1.$\!^{\prime\prime}$0--1.$\!^{\prime\prime}$5 southwest of the $^{12}$CO ($J$ = 3--2) peak, as defined by \cite{vit14}. The comparison of these molecular abundances in each knot is discussed in Section~\ref{sec:dis}.

An elongated structure toward the northeast can be seen in C$_{2}$H, HCN, HCO$^{+}$, HNC, CS, and CN. This feature is also observed in the $^{12}$CO distribution. This structure connects the CND to the surrounding SBR. In addition, we found strong spot emissions approximately 5$^{\prime\prime}$--6$^{\prime\prime}$ north and 3$^{\prime\prime}$ east of the AGN position along the central axis of the northeastern outflow for many molecules such as HC$^{18}$O$^{+}$, c-C$_{3}$H$_{2}$, HC$^{15}$N, HN$^{13}$C, HNCO, CH$_{3}$CN, C$^{34}$S, and CS. This position coincides with the hot spot of the AGN jet detected with the Very Large Array (VLA) at a wavelength of 6 cm \citep{gal96}. \cite{das06} proposed that there is an interaction between gas in the host galaxy and ionizing radiation in a part of the northeastern outflow. Therefore, these bright spots likely appear because of gas interactions. This position coincides with the bow-shock arc observed in the cold dust map reported by \cite{gar14}. However, distinctive structures in all the molecules were not found in the southwestern outflow region.

The amount of missing flux for each molecular line is summarized in the third and fourth column of Table~\ref{tab:line} as recoverable flux compared with the total fluxes obtained with two single-dish telescopes. The integrated fluxes obtained with the NRO 45-m telescope ($\theta_{\rm mb}$ = 15.$\!^{\prime\prime}$2--19.$\!^{\prime\prime}$1) and IRAM 30-m telescope ($\theta_{\rm mb}$ = 21$\!^{\prime\prime}$--29$\!^{\prime\prime}$) were obtained from \cite{tak19} and \cite{ala15}. The fluxes were recovered by approximately 60--80\% for most of the molecules detected in this line survey with the ALMA 12-m antennae. However, several molecules, such as HC$_{3}$N ($J$ = 10--9) and N$_{2}$H$^{+}$, show recovered fluxes below 40\% relative to both single-dish telescopes. The reason for this large missing flux of HC$_{3}$N ($J$ = 10--9) could be explained by the fact that the flux obtained for this molecule with the NRO 45-m telescope was not accurate owing to the large noise level and baseline fluctuation \citep[see][]{nak18, tak19}. For N$_{2}$H$^{+}$, the reason for the large missing flux is the maximum recoverable scale of the array configuration (see Section~\ref{sec:obs}).

\subsection{Line properties at 350 pc scale}

\begin{figure*}
\plotone{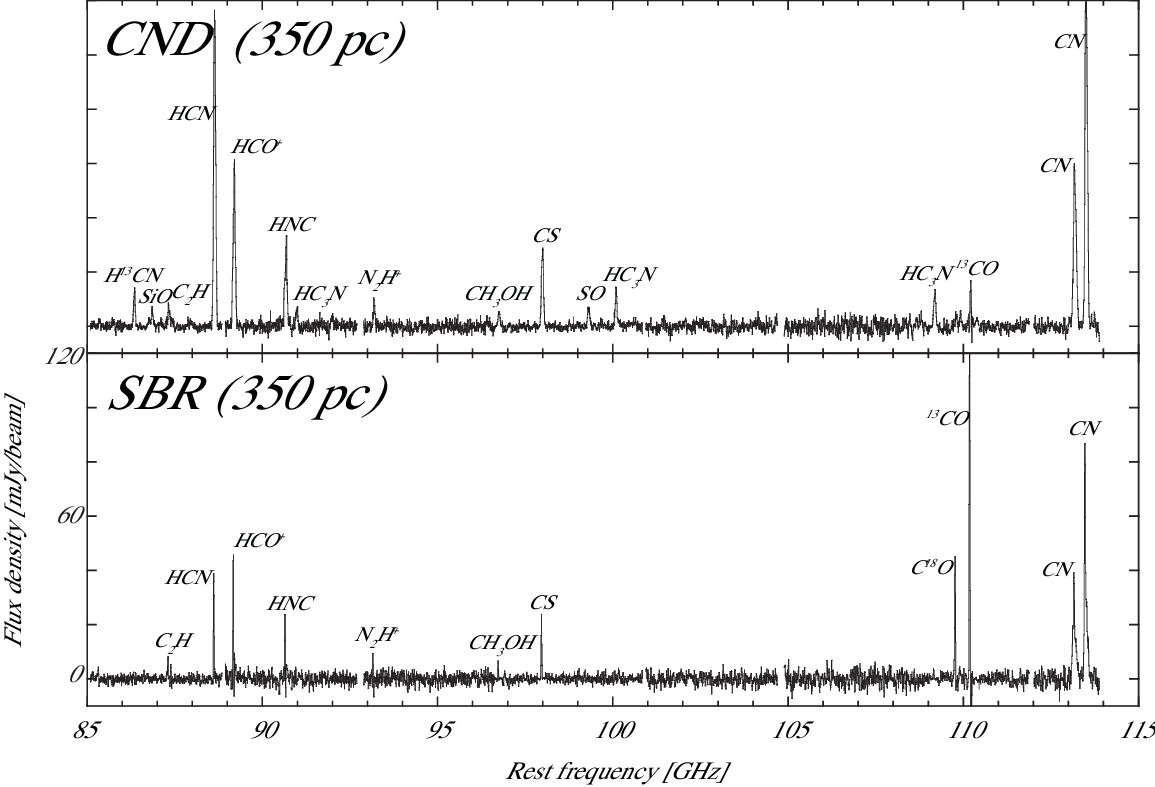}
\caption{The obtained spectra with spatial resolution of 5.$\!^{\prime\prime}$0 ($\sim$350 pc) from 85 to 114 GHz in the CND (upper panel) and the SBR (lower panel). The vertical and horizontal axes are the beam-averaged flux density and rest frequency, respectively. The velocity resolution is 20 km s$^{-1}$, and the primary beam correction is applied. The names of major molecular line carriers are indicated above the spectra.
\label{fig:350all}}
\end{figure*}

\begin{figure*}
\plotone{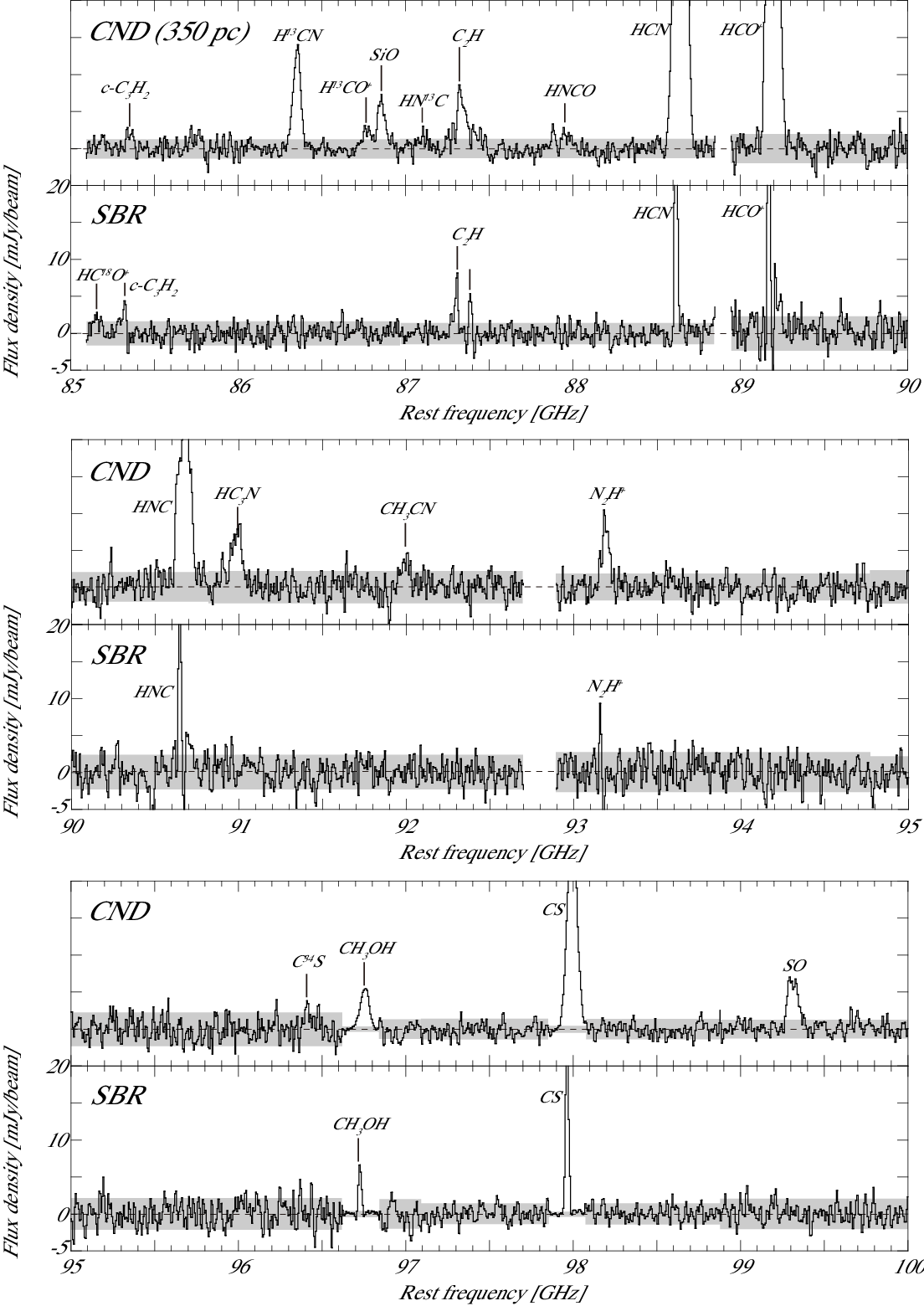}
\caption{Enlarged spectra with a spatial resolution of 5.$\!^{\prime\prime}$0 ($\sim$350 pc) from 85 to 100 GHz in the CND (upper panel) and the SBR (lower panel). The vertical and horizontal axes are the beam-averaged flux density and rest frequency, respectively. The velocity resolution is 20 km s$^{-1}$, and the primary beam correction is applied. The shaded regions refer to the range of $\pm$1$\sigma$ noise level. The names of the molecular line carriers, which have a peak flux of more than 1$\sigma$, are indicated above the spectra.
\label{fig:350_1}}
\end{figure*}

\begin{figure*}
\plotone{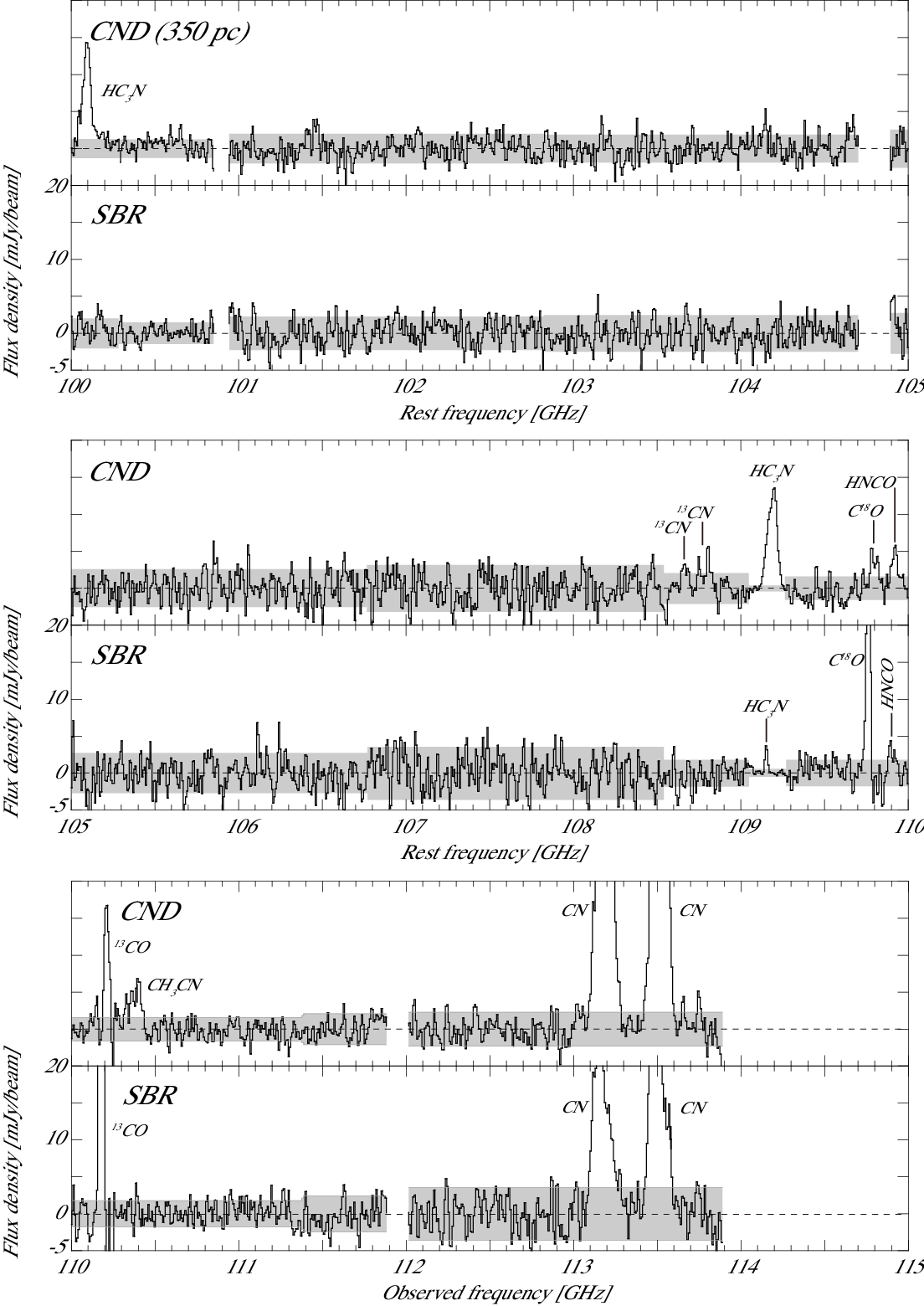}
\caption{Enlarged spectra with a spatial resolution of 5.$\!^{\prime\prime}$0 ($\sim$350 pc) from 100 to 114 GHz in the CND (upper panel) and the SBR (lower panel).
\label{fig:350_2}}
\end{figure*}

To investigate the differences in the physical/chemical properties of the interstellar medium affected by the AGN and starburst sources, we extracted the line spectra in the AGN and SBR positions with a convolved beam size of 5.$\!^{\prime\prime}$0 ($\sim$350 pc). This is the highest matched angular resolution in this line survey (see Table~\ref{tab:rec}). As shown in Figure~\ref{fig:beam}(a), the beam in the AGN position covers the entire structure of the CND, which was covered by an asymmetric elongated ring of 4$^{\prime\prime}$ $\times$ 2.$\!^{\prime\prime}$8, as indicated by \citet{gar14}.

The spectra from 85 to 114 GHz at these two positions are shown in Figure~\ref{fig:350all}. The velocity resolution was set to 20 km s$^{-1}$, and primary beam correction was applied to all spectra (Figures~\ref{fig:350all}--\ref{fig:60_3}). To obtain information from the same interstellar gas component among all 24 SPWs (see Table~\ref{tab:rec}), the minimum baseline is standardized for imaging and for extracting spectra. The $uv$ range is used for more than 15 k$\lambda$, which is the largest common minimum $uv$ range of all SPWs (see Table~\ref{tab:rec}).

While more than 15 species, such as HCN, HCO$^{+}$, CO isotopologues, and CN, were detected in both the CND and SBR, the flux density ratios were different between these regions. For example, $R_{{\rm HCN/HCO^{+}}}$ is approximately two at the CND, whereas it is below unity for the SBR. The flux ratios of $R_{{\rm H^{13}CN/CS}}$, $R_{{\rm SiO/CS}}$, and $R_{{\rm HC_{3}N/CS}}$ at the CND (0.4, 0.2, and 0.3--0.4) are significantly larger than those in the SBR ($<$0.07, $<$0.07, and 0.1), despite the fact that the peak fluxes of CS are almost similar at the CND and SBR. The lines of the CO isotopologues are very weak in the CND.

Figures~\ref{fig:350_1} and~\ref{fig:350_2} show the detailed spectra from 85 to 114 GHz at 5 GHz intervals. Molecular line identification is only performed for peak fluxes above the 1$\sigma$ noise level (shaded regions in the images) by eye, and the results are displayed with the names of the molecular carriers of lines. The integrated fluxes of each molecule are shown in the 5th and 6th columns of Table~\ref{tab:line}, and the non-detected lines (peak flux $<$1$\sigma$) are shown as the upper limit.

\subsection{Line properties at 60 pc scale} \label{sec:60}

\begin{figure*}
\plotone{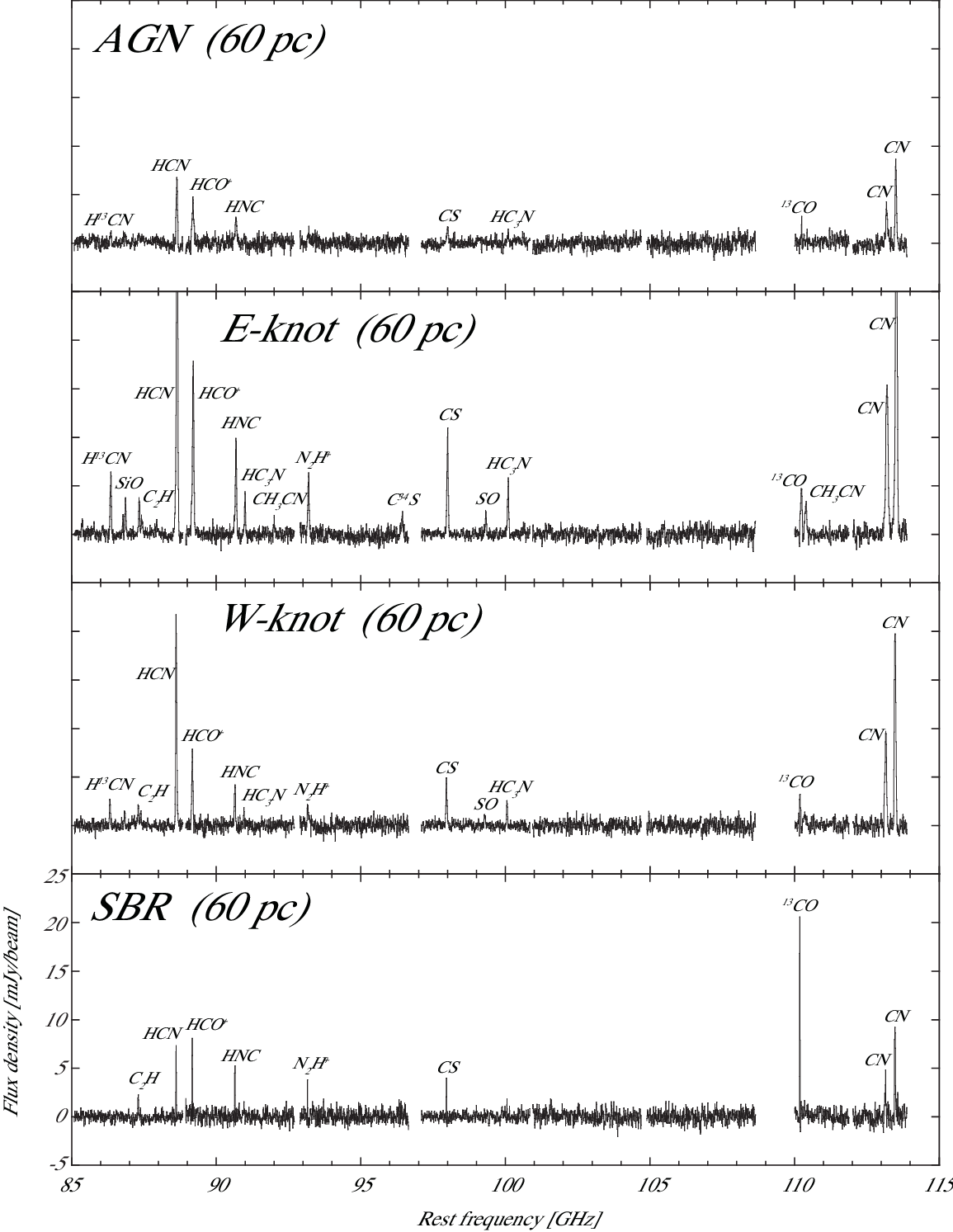}
\caption{Obtained spectra with a spatial resolution of 0.$\!^{\prime\prime}$9 ($\sim$60 pc) from 85 to 114 GHz in the AGN, the E-knot, the W-knot, and the SBR (from top down). The vertical and horizontal axes are the beam-averaged flux density and rest frequency, respectively. The velocity resolution is 20 km s$^{-1}$, and the primary beam correction is applied. The names of major molecular line carriers are indicated above the spectra.
\label{fig:60pcall}}
\end{figure*}

\begin{figure*}
\plotone{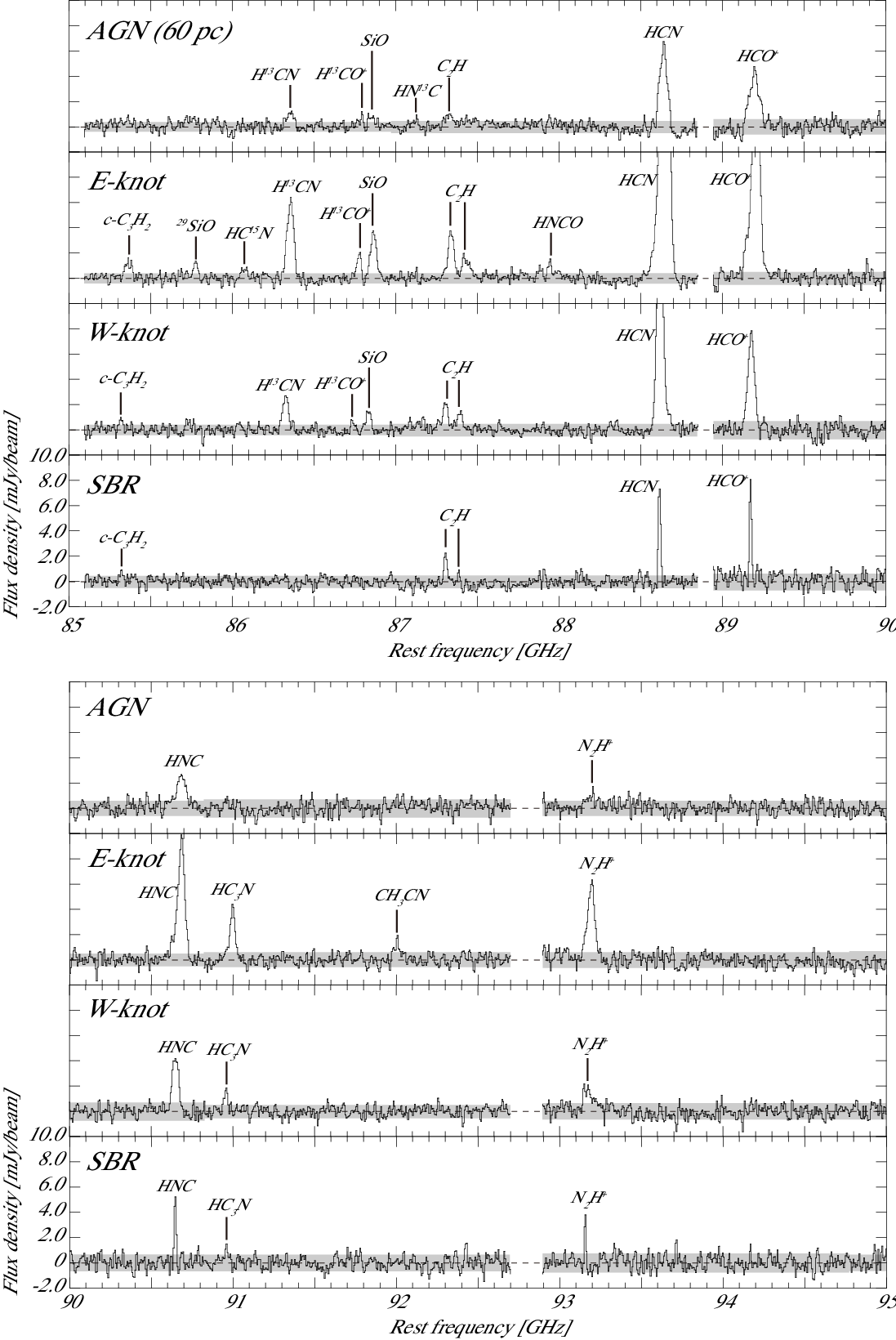}
\caption{Enlarged spectra with a spatial resolution of 0.$\!^{\prime\prime}$9 ($\sim$60 pc) from 85 to 95 GHz in four regions (see the caption of Figures~\ref{fig:60pcall}). The vertical and horizontal axes are the beam-averaged flux density and rest frequency, respectively. The velocity resolution is 20 km s$^{-1}$, and the primary beam correction is applied. The shaded regions represent the range of the $\pm$1$\sigma$ noise level. The names of the molecular line carriers, for which the peak flux is more than 1$\sigma$, are indicated above the spectra.
\label{fig:60_1}}
\end{figure*}

\begin{figure*}
\plotone{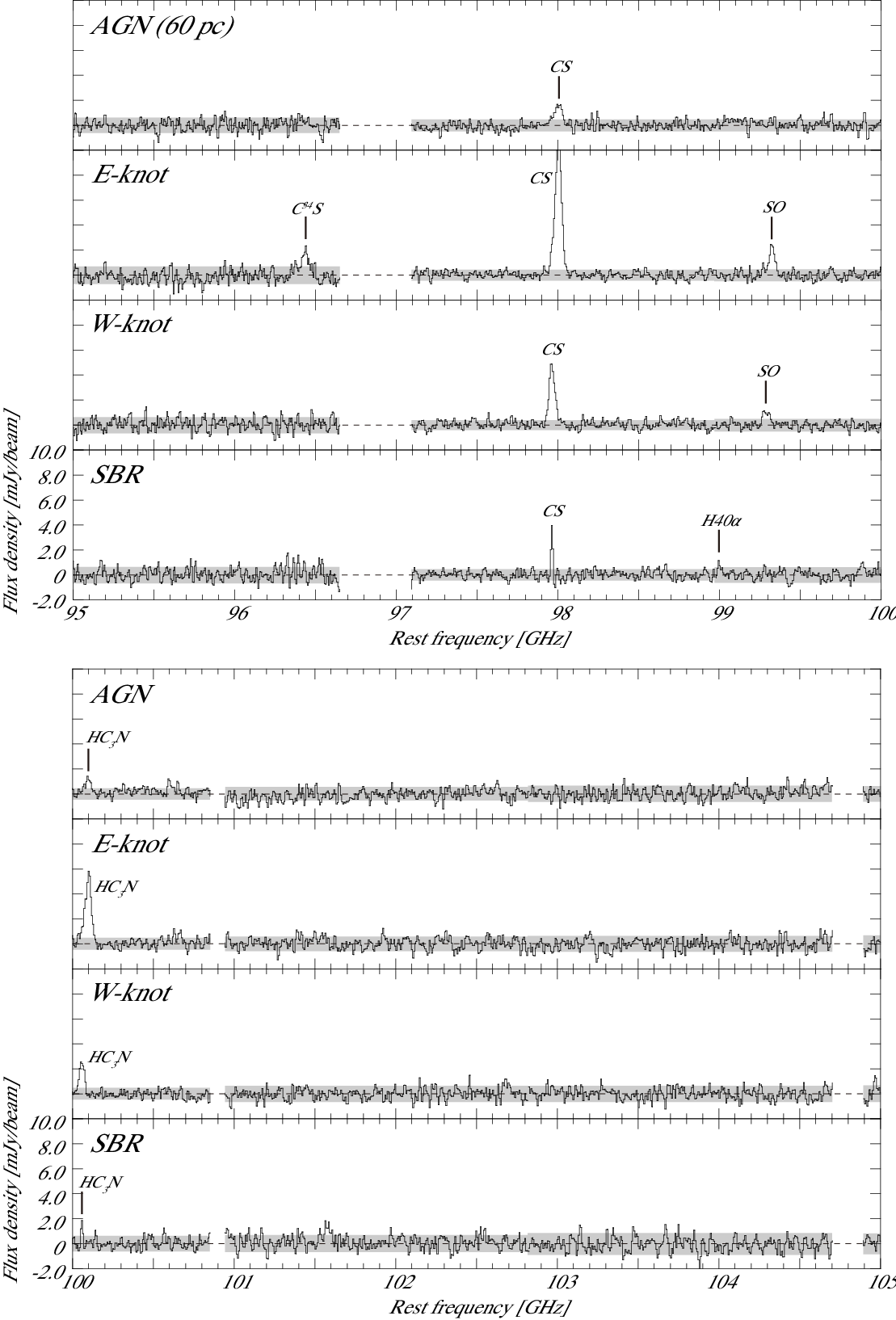}
\caption{Enlarged spectra with a spatial resolution of 0.$\!^{\prime\prime}$9 ($\sim$60 pc) from 95 to 105 GHz in four regions (see the caption of Figures~\ref{fig:60pcall}).
\label{fig:60_2}}
\end{figure*}

\begin{figure*}
\plotone{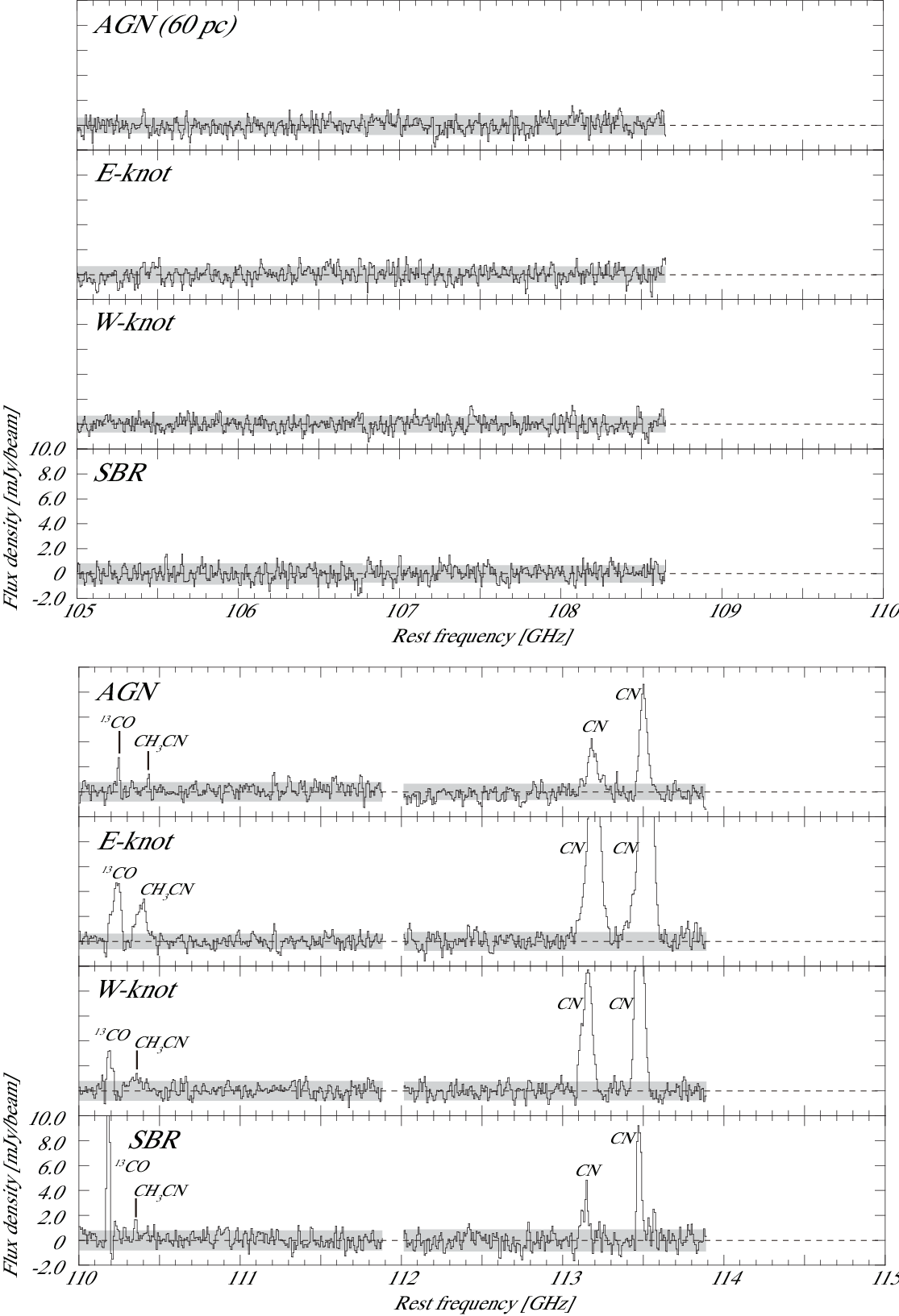}
\caption{Enlarged spectra with a spatial resolution of 0.$\!^{\prime\prime}$9 ($\sim$60 pc) from 105 to 114 GHz in four regions (see the caption of Figures~\ref{fig:60pcall}).
\label{fig:60_3}}
\end{figure*}

To determine the difference in physical/chemical properties within the CND, that is, the central position (AGN position), E-knot, and W-knot, we extracted a spectrum from the data cubes at these positions as well as the SBR with a convolved beam size of 0.$\!^{\prime\prime}$9 ($\sim$60 pc). This is the lowest common value among the observations of the extended antenna configuration (greater than 1600 m; Table~\ref{tab:obs}). As shown in Figure~\ref{fig:beam}(b), the beams at the E-knot, AGN, and W-knot positions do not overlap. However, the diffuse and extended molecular gas components around the knots likely contaminate the spectrum at the AGN position. We believe that our AGN spectrum is more or less contaminated, especially by the neighboring E-knot. Thus, our AGN spectrum does not originate solely from the AGN torus \citep{gar16, ima18, ima20}. In such a case, the source size is smaller than the convolved beam size; thus, the beam-averaged integrated flux and the column density may be underestimated. Because we would like to obtain the line properties under the same conditions at all sampling positions without arbitrary assumptions, the beam-filling factor is assumed to be unity.

The spectra from 85 to 114 GHz at these four positions are shown in Figure~\ref{fig:60pcall}. The velocity resolution was set to 20 km s$^{-1}$, and primary beam correction was applied for all images. The $uv$ range is used for more than 15 k$\lambda$, which is the same as that in the previous section. Figures~\ref{fig:60_1}, ~\ref{fig:60_2}, and ~\ref{fig:60_3} show the detailed spectra from 85 GHz to 114 GHz at 5 GHz intervals. The integrated fluxes of the detected lines are shown from the 7th to 10th columns of Table~\ref{tab:line}, and the non-detected lines (peak flux $<$1$\sigma$) are shown in the upper limit.

\begin{deluxetable*}{clcccccccccc}
\tablenum{3}
\tablecaption{Recovered and integrated flux for each molecular line.\label{tab:line}}
\tablewidth{0pt}
\tablehead{
\colhead{Frequency$^{a}$} & \colhead{Molecule} & \multicolumn2c{Recovery} & & \multicolumn7c{$\int$flux$dv$ (Jy beam$^{-1}$ km s$^{-1}$)} \\
\colhead{(GHz)} & \colhead{(Transition)} &  \multicolumn2c{(\%)} & & \multicolumn2c{5.$\!^{\prime\prime}$0 resolution} & & \multicolumn4c{0.$\!^{\prime\prime}$9 resolution} \\
\cline{3-4}
\cline{6-7}
\cline{9-12}
 & & \colhead{NRO$^{b}$} & \colhead{IRAM$^{c}$} & & \colhead{CND} & \colhead{SBR} & & \colhead{AGN} & \colhead{E-knot} & \colhead{W-knot} & \colhead{SBR}
}
\startdata
85.162223 & HC$^{18}$O$^{+}$ (1--0) & ---  & --- & & 0.23$\pm$0.09$^{d}$ & 0.36$\pm$0.11 & & $<$0.02 & $<$0.02 & $<$0.03 & $<$0.03 \\
85.338894 & c-C$_{3}$H$_{2}$ (2$_{1,2}$--1$_{0,1}$) & 80$\pm$17 & --- & & 0.40$\pm$0.09 & 0.51$\pm$0.11 & & 0.09$\pm$0.02 & 0.22$\pm$0.03 & 0.09$\pm$0.02 & 0.08$\pm$0.03$^{d}$ \\
85.759194 & $^{29}$SiO (2$_{0}$--1$_{0}$) & --- & --- & & 0.33$\pm$0.08 & $<$0.11 & & 0.10$\pm$0.02 & 0.15$\pm$0.02 & 0.08$\pm$0.02 & $<$0.03 \\
86.054966 & HC$^{15}$N (1--0) & --- & --- & & $<$0.1 & $<$0.11 & & $<$0.03 & 0.13$\pm$0.03 & 0.07$\pm$0.02 & $<$0.03 \\
86.339921 & H$^{13}$CN (1--0) & 88$\pm$17 & 62$\pm$10 & & 2.67$\pm$0.13 & $<$0.11 & & 0.22$\pm$0.03 & 1.06$\pm$0.03 & 0.40$\pm$0.03 & $<$0.03 \\
86.754288 & H$^{13}$CO$^{+}$ (1--0) & --- & --- & & 0.53$\pm$0.1 & $<$0.11 & & 0.11$\pm$0.02 & 0.22$\pm$0.02 & 0.07$\pm$0.03$^{d}$ & $<$0.03 \\
86.846985 & SiO (2--1) & 108$\pm$31 & 52$\pm$10 & & 1.40$\pm$0.12 & $<$0.11 & & 0.16$\pm$0.03 & 0.58$\pm$0.03 & 0.2$\pm$0.03 & $<$0.03 \\
87.090825 & HN$^{13}$C (1--0) & --- & --- & & 0.28$\pm$0.09 & $<$0.09 & & 0.08$\pm$0.03$^{d}$ & $<$0.03 & 0.08$\pm$0.03$^{d}$ & $<$0.03 \\
87.316898 & C$_{2}$H (1--0) & 138$\pm$9 & 54$\pm$2 & & 2.14$\pm$0.14 & 1.17$\pm$0.13 & & 0.41$\pm$0.06 & 0.92$\pm$0.05 & 0.55$\pm$0.06 & 0.22$\pm$0.04 \\
87.925237 & HNCO (4$_{0,4}$--3$_{0,3}$) & 45$\pm$12 & 56$\pm$89 & & 0.73$\pm$0.13 & 0.14$\pm$0.09$^{d}$ & & 0.17$\pm$0.03 & 0.23$\pm$0.03 & $<$0.04 & 0.1$\pm$0.03 \\
88.631602 & HCN (1--0) & 108$\pm$3 & 60$\pm$1 & & 28.04$\pm$0.14 & 3.05$\pm$0.09 & & 1.19$\pm$0.04 & 8.42$\pm$0.05 & 3.31$\pm$0.05 & 0.45$\pm$0.02 \\
89.188525 & HCO$^{+}$ (1--0) & 118$\pm$6 & 67$\pm$1 & & 14.0$\pm$0.18 & 2.75$\pm$0.11 & & 1.19$\pm$0.06 & 3.69$\pm$0.06 & 1.27$\pm$0.06 & 0.45$\pm$0.03 \\
90.663568 & HNC (1--0) & 147$\pm$12 & 63$\pm$3 & & 7.39$\pm$0.19 & 1.65$\pm$0.11 & & 0.73$\pm$0.07 & 1.77$\pm$0.05 & 0.68$\pm$0.06 & 0.26$\pm$0.03 \\
90.979023 & HC$_{3}$N (10--9) & 13$\pm$4 & 50$\pm$17 & & 2.20$\pm$0.23 & 0.23$\pm$0.09$^{d}$ & & 0.18$\pm$0.06$^{d}$ & 0.75$\pm$0.06 & 0.21$\pm$0.04 & 0.09$\pm$0.03$^{d}$ \\
91.987088 & CH$_{3}$CN (5$_{K}$--4$_{K}$) & --- & --- & & 0.84$\pm$0.17 & $<$0.15 & & $<$0.06 & 0.25$\pm$0.05 & $<$0.04 & $<$0.04 \\
93.173977 & N$_{2}$H$^{+}$ (1--0) & 24$\pm$5 & 17$\pm$1 & & 1.55$\pm$0.16 & 0.40$\pm$0.24$^{d}$ & & 0.26$\pm$0.05 & 1.10$\pm$0.06 & 0.43$\pm$0.06 & 0.15$\pm$0.03 \\
96.412950 & C$^{34}$S (2--1) & --- & --- & & 0.39$\pm$0.14$^{d}$ & $<$0.13 & & $<$0.05 & 0.45$\pm$0.07 & $<$0.05 & $<$0.04 \\
96.744550 & CH$_{3}$OH (2$_{K}$--1$_{K}$) & 37$\pm$7 & 100$\pm$17 & & 1.12$\pm$0.04 & 0.44$\pm$0.04 & & & & & \\
97.980953 & CS (2--1) & 56$\pm$2 & 81$\pm$2 & & 6.78$\pm$0.04 & 1.68$\pm$0.03 & & & & & \\
 & & 83$\pm$5 & 60$\pm$3 & & & & & 0.32$\pm$0.04 & 1.74$\pm$0.04 & 0.70$\pm$0.04 & 0.17$\pm$0.02 \\
99.02296 & H40$\alpha$ & --- & --- & & $<$0.08 & $<$0.13 & & $<$0.03 & $<$0.03 & $<$0.03 & 0.08$\pm$0.03$^{d}$ \\
99.299870 & SO (3$_{2}$--2$_{1}$) & 84$\pm$21 & 84$\pm$12 & & 1.51$\pm$0.12 & $<$0.18 & & & & & \\
 & & 60$\pm$26 & 41$\pm$17 & & & & & $<$0.04 & 0.36$\pm$0.04 & 0.17$\pm$0.03 & $<$0.04 \\
100.076385 & HC$_{3}$N (11--10) & 75$\pm$14 & 87$\pm$6 & & 2.88$\pm$0.12 & 0.24$\pm$0.18$^{d}$ & & & & & \\
 & & 81$\pm$22 & 66$\pm$14 & & & & & 0.17$\pm$0.03 & 0.84$\pm$0.04 & 0.29$\pm$0.03 & 0.08$\pm$0.02 \\
108.651297 & $^{13}$CN (1--0) & 58$\pm$9 & --- & & 0.65$\pm$0.19 & $<$0.16 & & & & & \\
109.173638 & HC$_{3}$N (12--11) & 61$\pm$5 & 156$\pm$8 & & 2.49$\pm$0.03 & 0.21$\pm$0.05 & & & & & \\
109.782173 & C$^{18}$O (1--0) & 44$\pm$3 & 55$\pm$2 & & 0.65$\pm$0.14 & 3.32$\pm$0.16 & & & & & \\
109.905753 & HNCO (5$_{0,5}$--4$_{0,4}$) & 43$\pm$8 & --- & & 0.72$\pm$0.14 & 0.27$\pm$0.16$^{d}$ & & & & & \\
110.201354 & $^{13}$CO (1--0) & 61$\pm$1 & 61$\pm$1 & & 1.67$\pm$0.14 & 9.09$\pm$0.16 & & & & & \\
 & & 32$\pm$1 & 61$\pm$2 & & & & & 0.17$\pm$0.04 & 0.79$\pm$0.04 & 0.38$\pm$0.06 & 1.06$\pm$0.04 \\
110.383500 & CH$_{3}$CN (6$_{K}$--5$_{K}$) & 115$\pm$23 & 49$\pm$35 & & 1.11$\pm$0.16 & $<$0.16 & & & & & \\
 & & 54$\pm$26 & 23$\pm$29 & & & & & 0.10$\pm$0.04$^{d}$ & 0.64$\pm$0.05 & 0.26$\pm$0.06 & 0.11$\pm$0.03 \\
113.191279 & CN (1--0) & 141$\pm$4 & 103$\pm$3 & & 14.89$\pm$0.21 & 8.49$\pm$0.32 & & 0.79$\pm$0.06 & 4.06$\pm$0.08 & 1.87$\pm$0.06 & 0.47$\pm$0.06 \\
113.490970 & CN (1--0) & 154$\pm$3 & 87$\pm$1 & & 27.57$\pm$0.21 & 13.09$\pm$0.32 & & 1.24$\pm$0.05 & 7.85$\pm$0.09 & 3.29$\pm$0.06 & 0.85$\pm$0.05 \\
\enddata
\tablecomments{$^{a}$ Rest frequencies are referenced in \cite{lil68} for the recombination line and \cite{lov04} for others. $^{b}$ Recovery fluxes with the NRO 45-m telescope were calculated based on \cite{tak19}. $^{c}$ Recovery fluxes with the IRAM 30-m telescope were calculated based on \cite{ala15}. $^{d}$ Low S/N ($<$3$\sigma$).}
\end{deluxetable*}

\subsection{Estimation of column densities}

\begin{deluxetable*}{lcccccccccc}
\tablenum{4}
\tablecaption{Column densities of each molecule.\label{tab:nmol}}
\tablewidth{0pt}
\tablehead{
\colhead{Molecule} & \multicolumn2c{$T_{\rm rot}$ (K)} & \multicolumn7c{$N_{\rm mol}$ (cm$^{-2}$)} \\
 & & & & \multicolumn2c{5.$\!^{\prime\prime}$0 resolution} & & \multicolumn4c{0.$\!^{\prime\prime}$9 resolution} \\
\cline{2-3}
\cline{5-6}
\cline{8-11}
 & \colhead{CND} & \colhead{SBR} & & \colhead{CND} & \colhead{SBR} & & \colhead{AGN} & \colhead{E-knot} & \colhead{W-knot} & \colhead{SBR}
}
\startdata
HC$^{18}$O$^{+}$ & 15$\pm$5 & 10$\pm$5 & & 2.0$^{+1.4}_{-1.1}$(12) & 2.5$^{+1.7}_{-1.1}$(12) & & $<$5.4(12) & $<$5.4(12) & $<$8.1(12) & $<$6.4(12) \\
cyclic-C$_{3}$H$_{2}$ & 15$\pm$5 & 10$\pm$5 & & 2.4$^{+1.7}_{-1.1}$(13) & 2.1$^{+1.7}_{-1.0}$(13) & & 1.7$^{+1.2}_{-0.8}$(14) & 4.1$^{+2.3}_{-1.7}$(14) & 1.7$^{+1.2}_{-0.8}$(14) & 1.0$^{+1.1}_{-0.6}$(14) \\
$^{29}$SiO & 15$\pm$5 & 10$\pm$5 & & 4.6$^{+2.4}_{-1.9}$(12) & $<$1.2(12) & & 4.2$^{+2.0}_{-1.6}$(13) & 6.5$^{+2.6}_{-2.1}$(13) & 3.7$^{+1.9}_{-1.5}$(13) & $<$1.0(13) \\
HC$^{15}$N & 15$\pm$5 & 10$\pm$5 & & $<$1.3(12) & $<$1.3(12) & & $<$1.4(13) & 5.6$^{+3.0}_{-2.3}$(13) & 3.0$^{+1.8}_{-1.4}$(13) & $<$1.1(13) \\
H$^{13}$CN & 15$\pm$5 & 10$\pm$5 & & 3.8$^{+1.1}_{-1.0}$(13) & $<$1.2(12) & & 9.7$^{+3.8}_{-3.1}$(13) & 4.7$^{+1.2}_{-1.1}$(14) & 1.7$^{+0.6}_{-0.5}$(14) & $<$1.0(13) \\
H$^{13}$CO$^{+}$ & 15$\pm$5 & 10$\pm$5 & & 4.4$^{+2.0}_{-1.6}$(12) & $<$7.1(11) & & 2.7$^{+1.3}_{-1.0}$(13) & 5.6$^{+1.9}_{-1.6}$(13) & 1.8$^{+1.4}_{-1.0}$(13) & $<$6.0(12) \\
SiO & 15$\pm$5 & 10$\pm$5 & & 1.9$^{+0.6}_{-0.5}$(13) & $<$1.2(12) & & 6.7$^{+3.0}_{-2.4}$(13) & 2.4$^{+0.7}_{-0.6}$(14) & 8.2$^{+3.4}_{-2.7}$(13) & $<$9.8(12) \\
HN$^{13}$C & 15$\pm$5 & 10$\pm$5 & & 3.7$^{+2.3}_{-1.7}$(12) & $<$9.3(11) & & 3.2$^{+2.2}_{-1.6}$(13) & $<$1.2(13) & 3.5$^{+2.3}_{-1.7}$(13) & $<$9.6(12) \\
C$_{2}$H & 15$\pm$5 & 10$\pm$5 & & 4.4$^{+1.4}_{-1.2}$(14) & 1.9$^{+0.8}_{-0.5}$(14) & & 2.6$^{+1.1}_{-0.9}$(15) & 5.8$^{+1.7}_{-1.5}$(15) & 3.5$^{+1.3}_{-1.0}$(15) & 1.1$^{+0.6}_{-0.4}$(15) \\
HNCO & 9.9$^{+0.1}_{-5.8}$ & 10$\pm$5 & & 1.0$^{+0.1}_{-0.6}$(14) & 1.3$^{+3.5}_{-0.8}$(13) & & 3.6$^{+4.3}_{-0.6}$(14) & 4.9$^{+5.4}_{-0.6}$(14) & $<$8.4(13)& 2.0$\pm$0.6(14) \\
HCN & 15$\pm$5 & 10$\pm$5 & & 3.6$\pm$0.8(14) & 3.1$^{+0.8}_{-0.6}$(13) & & 4.7$^{+1.3}_{-1.1}$(14) & 3.4$^{+0.8}_{-0.7}$(15) & 1.3$\pm$0.3(15) & 1.4$^{+0.4}_{-0.3}$(14) \\
HCO$^{+}$ & 15$\pm$5 & 10$\pm$5 & & 1.0$^{+0.3}_{-0.2}$(14) & 1.5$^{+0.5}_{-0.4}$(13) & & 2.3$^{+0.7}_{-0.6}$(14) & 8.1$^{+2.2}_{-2.0}$(14) & 2.8$^{+0.9}_{-0.8}$(14) & 7.5$^{+2.3}_{-1.9}$(13) \\
HNC & 15$\pm$5 & 10$\pm$5 & & 8.4$^{+2.2}_{-1.9}$(13) & 1.5$^{+0.4}_{-0.3}$(13) & & 2.6$^{+0.9}_{-0.7}$(14) & 6.2$^{+1.6}_{-1.5}$(14) & 2.4$^{+0.8}_{-0.7}$(14) & 7.3$^{+1.9}_{-1.3}$(13) \\
HC$_{3}$N & 16.7$^{+3.9}_{-3.1}$ & 10$\pm$5 & & 5.9$^{+1.5}_{-0.8}$(13) & 0.9$^{+6.0}_{-0.4}$(13) & & 1.4$^{+0.7}_{-0.5}$(14) & 5.9$^{+1.3}_{-0.8}$(14) & 1.6$^{+0.6}_{-0.4}$(14) & 0.9$^{+2.0}_{-0.3}$(14) \\
CH$_{3}$CN & 11.5$^{+0.7}_{-0.2}$ & 10$\pm$5 & & 5.5$^{+4.4}_{-1.5}$(12) & $<$1.2(12) & & 1.9$\pm$0.8(13) & 1.2$^{+0.0}_{-0.1}$(14) & 5.0$\pm$0.1(13) & 2.2$\pm$0.6(13) \\
N$_{2}$H$^{+}$ & 15$\pm$5 & 10$\pm$5 & & 1.3$\pm$0.4(13) & 2.6$^{+0.7}_{-0.4}$(12) & & 6.6$^{+3.0}_{-2.4}$(13) & 2.8$^{+0.8}_{-0.7}$(14) & 1.1$\pm$0.4(14) & 3.1$^{+0.8}_{-0.5}$(13) \\
C$^{34}$S & 15$\pm$5 & 10$\pm$5 & & 8.7$^{+5.8}_{-4.3}$(12) & $<$2.3(12) & & $<$3.5(13) & 3.1$^{+1.3}_{-1.0}$(14) & $<$3.5(13) & $<$2.2(13) \\
CH$_{3}$OH & 15$\pm$5 & 10$\pm$5 & & 2.7$^{+1.1}_{-1.7}$(14) & 7.6$^{+6.4}_{-5.0}$(13) & & & & & \\
CS & 12.6$\pm$0.2 & 10.5$^{+0.8}_{-0.6}$ & & 1.4$\pm$0.03(14) & 3.0$^{+0.7}_{-0.3}$(13) & & 2.0$\pm$0.3(14) & 1.1$\pm$0.03(15) & 4.4$\pm$0.3(14) & 9.7$^{+2.2}_{-0.9}$(13) \\
SO & 15$\pm$5 & 10$\pm$5 & & 1.0$^{+0.4}_{-0.3}$(14) & $<$9.1(12) & & $<$8.4(13) & 7.6$^{+2.9}_{-2.6}$(14) & 3.7$^{+1.7}_{-1.4}$(14) & $<$6.2(13) \\
$^{13}$CN & 15$\pm$5 & 10$\pm$5 & & 1.7$^{+0.9}_{-0.7}$(14) & $<$3.4(12) & & & & & \\
C$^{18}$O & 13.9$^{+3.1}_{-2.0}$ & 8.0$\pm$0.3 & & 2.8$^{+1.3}_{-1.0}$(15) & 1.1$^{+0.3}_{-0.1}$(16) & & & & & \\
$^{13}$CO & 16.7$^{+1.2}_{-1.0}$ & 8.5$\pm$0.1 & & 7.6$^{+0.4}_{-0.3}$(15) & 3.0$\pm$0.04(16) & & 5.5$^{+2.8}_{-1.8}$(16) & 2.6$^{+0.7}_{-0.4}$(17) & 1.3$^{+0.5}_{-0.3}$(17) & 2.1$\pm$0.02(17) \\
CN & 15$\pm$5 & 10$\pm$5 & & 9.5$^{+7.8}_{-1.8}$(14) & 4.0$^{+0.9}_{-0.3}$(14) & & 1.4$^{+0.4}_{-0.3}$(15) & 8.2$^{+1.8}_{-1.6}$(15) & 3.6$^{+0.8}_{-0.7}$(15) & 7.5$^{+1.7}_{-0.6}$(14) \\
\enddata
\end{deluxetable*}

The column densities ($N_{\rm mol}$) of each molecule were estimated under the assumption of local thermodynamic equilibrium (LTE), and the estimated results are listed in Table~\ref{tab:nmol}. In the estimation, the rotational temperatures ($T_{\rm rot}$) and column densities are calculated using the rotation diagram approach under the assumption that all lines are optically thin and that a single excitation temperature characterizes all transitions \citep[e.g.,][]{gol99}. The assumption of LTE, especially in CND, may not be correct because the environment of this region is expected to be significantly affected by strong emissions and/or outflow from the AGN. However, this assumption is the simplest and most commonly used method for estimating molecular abundance. In addition, it allows us a direct comparison with similar studies in the literature \citep[e.g.,][]{mar06, bay09, ala13}. 

Because almost all the detected molecules in this work only show a single transition in the 3-mm band, we assume that $T_{\rm rot}$ in the CND (also in the AGN and the E- and W-knots) and that in the SBR are 15$\pm$5 K and 10$\pm$5 K, respectively. These values were determined based on the average of the previous ALMA observations, and the temperatures in the CND and the SBR were significantly different \citep{nak15}. Derived averaged rotational temperatures among a few molecules with two line detections (Figure~\ref{fig:rotdia} in the Appendix) are 13.6 K and 9.0 K toward the CND and the SBR, respectively. Therefore, we believe that the assumed $T_{\rm rot}$ values in the CND and SBR are reasonable. Note that the $T_{\rm rot}$ is a lower limit estimate for kinetic temperature ($T_{\rm kin}$) in general, and the value of $T_{\rm rot}$ is significantly smaller than that of $T_{\rm kin}$ in many cases. In fact, several works derived $T_{\rm kin}$ in the CND with non-LTE analysis, and they derived at least a few hundred kelvins from CO, HCN, HCO$^{+}$, CS, SiO, and HNCO \citep{vit14, sco20, but22, hua22}.

The column densities of CS, HCN, and HCO$^{+}$ using non-LTE analysis have been reported in previous works. We compare these reported values and our estimations to validate the assumption of LTE analysis. For CS, Scourfield et al. (2020) reported $N$(CS) using RADEX \citep{van07} and UCLCHEM \citep{hol17} codes. As a result, $N$(CS) in the CND and SBR only from RADEX is consistently compared with our LTE results (see Table 5 in Scourfield et al. 2020). However, $N$(CS) in the SBR from both UCLCHEM and RADEX is larger than our results \citep[see Table 6 in][]{sco20}. For HCN and HCO$^{+}$, $N$(HCN) and $N$(HCO$^{+}$) in both the CND and SBR from RADEX are significantly large (more than one order of magnitude) relative to the LTE analysis \citep[see Table 6 in][]{but22}. If the analysis with the LTE assumption in this work underestimates the column densities, the possibilities of this reason are the effect of optical depth, especially HCN, described above, and a missing flux in this ALMA band-3 observation using only the 12-m main array.

The optical depth of some major molecules with the possibility of being optically thick using the single-dish telescope was reported in the previous research \citep[see Table 2 in][]{nak18}. Based on this result, $^{12}$CO, HCN, and CN are optically thick in the case of the beam size of $\sim$1 kpc. Note that $^{12}$CO is not included in this work. Thus, we estimate the optical depth of these molecules in ALMA observations. The optical depth of HCN is estimated using an assumption of a $^{12}$C/$^{13}$C ratio \citep[$\sim$38;][]{tan19}, and the values with 60 and 350 pc are 7.8 and 3.7, respectively. For CN, the value is estimated to be about 0.1 with both 60 and 350 pc using the intensity ratio of the hyperfine structure \citep{ska83}. Therefore, HCN is optically thick, and the column density is possibly underestimated. We do not use HCN but mainly H$^{13}$CN to discuss molecular abundance in this paper.

\begin{figure*}
\plotone{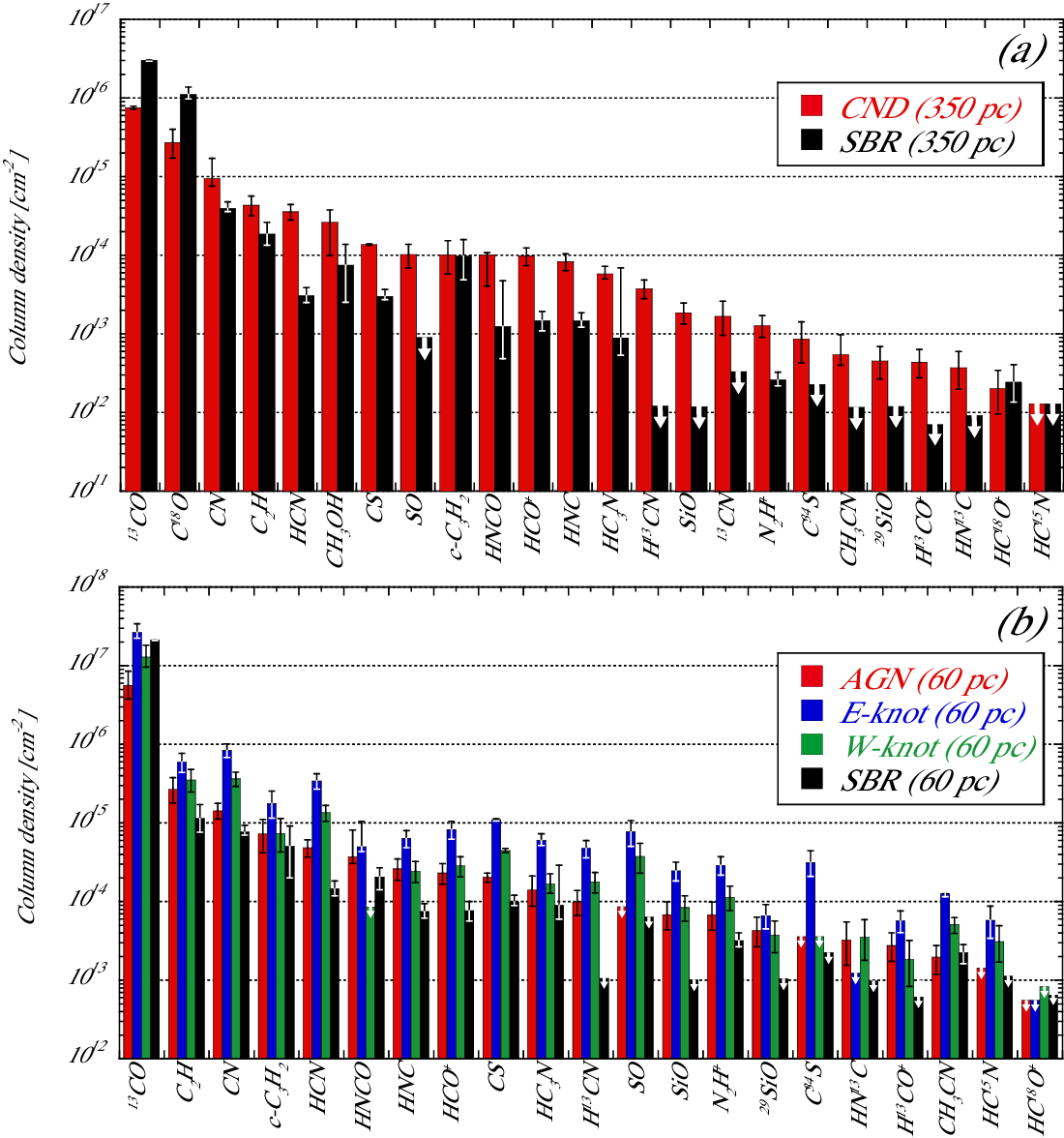}
\caption{Comparison of column densities of each molecule in (a) 350 pc scale and (b) 60 pc scale. The bar graphs in red, blue, green, and black are for the CND (or AGN), the E-knot, the W-knot, and the SBR, respectively. The order of molecules is arranged in descending order from left to right based on the column density toward CND in (a) or AGN in (b). Arrows indicate upper limits. \label{fig:colm}}
\end{figure*}

\section{Discussion} \label{sec:dis}

\begin{figure*}
\plotone{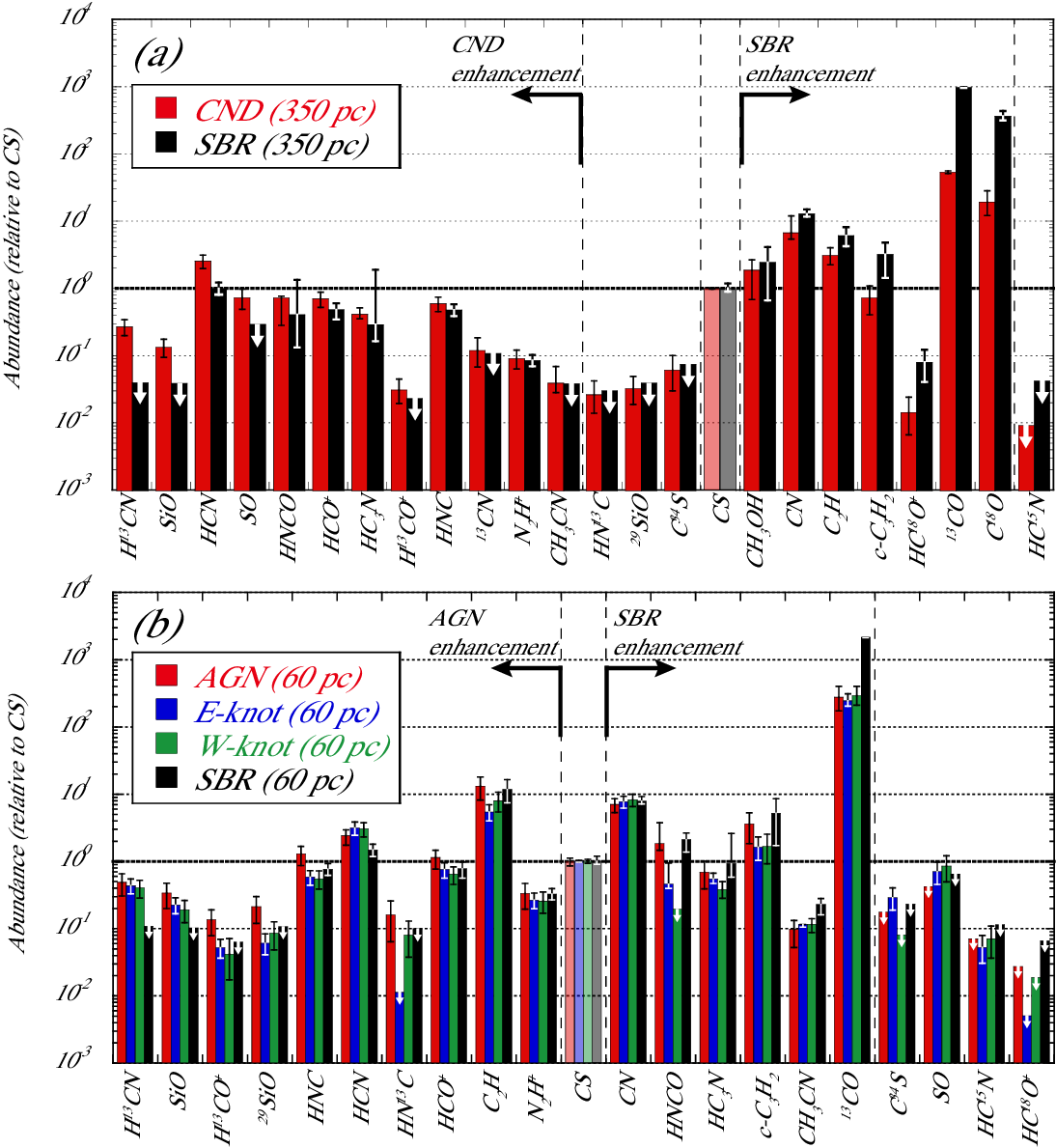}
\caption{Fractional abundances relative to CS in (a) the 350 pc scale and (b) the 60 pc scale. Ratios over unity, which is indicated by the bold dashed line, represents the enhancement of molecular abundances relative to CS. White arrows indicate the upper limits of the relative abundances. The order of the molecules is arranged in descending order from left to right based on the difference between the CND and the SBR for (a) and between the AGN and the SBR for (b). Therefore, the molecules on the left side of CS are enhancements in the CND (or AGN) compared with the SBR; in contrast, those on the right side of CS are enhancements in the SBR. Note that C$^{34}$S, SO, HC$^{15}$N, and HC$^{18}$O$^{+}$ are not detected in both the CND (or AGN) and SBR.
\label{fig:abun}}
\end{figure*}

\begin{figure*}
\plotone{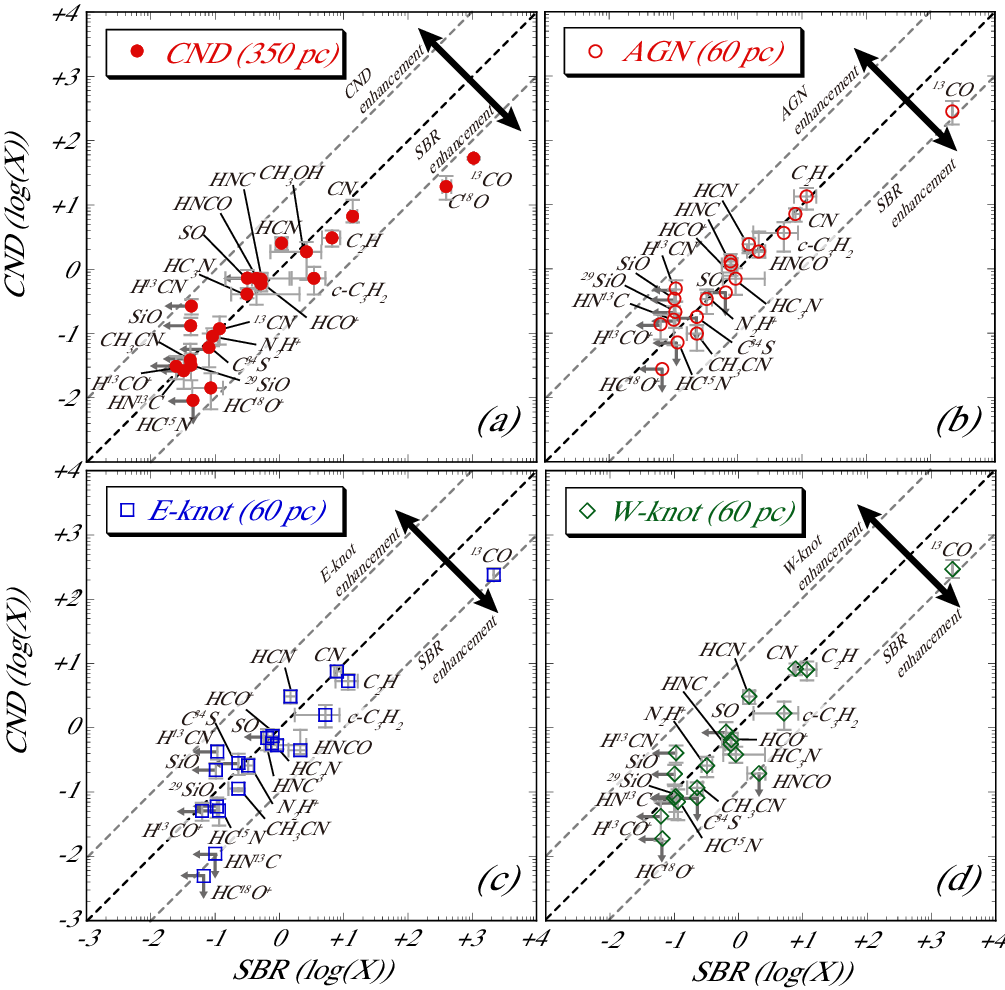}
\caption{Plots of the fractional abundances relative to CS (a) between CND and SBR with the 350 pc scale, (b) between AGN and SBR with the 60 pc scale, (c) between E-knot and SBR with the 60 pc scale, and (d) between W-knot and SBR with the 60 pc scale. These results represent the abundance correlation between the CND and the SBR. Plots above the bold dashed line represent the enhanced molecules and plots below represent deficient ones in the CND (or AGN).
\label{fig:corri}}
\end{figure*}

In this section, we discuss primarily the physical and chemical effects of the AGN on molecular composition and abundance. First, $N_{\rm mol}$ of each molecule and the fractional abundances relative to CS in the CND were compared with those in the SBR. Next, we investigated the difference between the fractional abundances for those obtained with ALMA (spatial resolutions of 60 pc and 350 pc) and the NRO 45-m single-dish telescope ($\sim$1.2 kpc) because it is expected that the higher the resolution, the more clear the effect of the AGN.

\subsection{Column densities in the CND vs. SBR}
Figure~\ref{fig:colm} shows the column density of each molecule with spatial resolutions of 350 pc and 60 pc in the CND and SBR. The order of molecules is arranged in descending order from left to right based on $N_{\rm mol}$ toward the CND (350 pc scale) in Figure~\ref{fig:colm}(a), and toward the AGN (60 pc scale) in (b). 

At the 350 pc scale, column densities in the CND are systematically higher than those in the SBR, except for $^{13}$CO, C$^{18}$O, and HC$^{18}$O$^{+}$. Note that the difference in HC$^{18}$O$^{+}$ between the CND and SBR is within the error bar. At the 60 pc scale, column densities in the AGN are systematically higher than those in the SBR, except for $^{13}$CO. This feature is consistent with the single-dish measurements reported in the literature \citep[e.g.,][]{nak18}. There are no molecules with a higher density in the AGN position than in the E- and/or W- knots. We believe that the dense molecular torus at the AGN position, whose density may be close to that of the knot regions reported by \cite{gar16} and \cite{ima18}, is not significant in this observation. In addition, one of the reasons for the lower column densities could be the effect of the beam-filling factor (see Section~\ref{sec:60}). None of the molecules detected in either knot has a higher $N_{\rm mol}$ in the W-knot than in the E-knot in the 60 pc scale except for HN$^{13}$C. The high value for HN$^{13}$C in the W-knot is not significant because the detection is judged with low-S/N (see Table~\ref{tab:line}). \cite{sai22a} reported the integrated intensity map of atomic carbon [C$\mathrm{I}$], and that in the W-knot is higher than that in the E-knot owing to the effect of the AGN jet and outflow. In this work, it is difficult to identify this effect because the samples are only at three positions, and these are too discrete. In a recent study, \cite{sai22b} also discussed the effect on molecules using a principal component analysis.

\subsection{Fractional abundances relative to CS}
Fractional abundances in the CND and SBR at 350 pc and 60 pc resolution are shown in Figure~\ref{fig:abun} as a column graph, and Figure~\ref{fig:corri} as a scatter plot. The molecule used for the normalization of abundance was CS, which is one of the best tracers of dense gas because of its high critical density. \cite{sco20} described the chemistry and excitation in individual regions located in the CND and the SBR in NGC 1068 based on the multi-transitions of CS with ALMA. In addition, fractional abundance with respect to CS is useful for comparing different chemical compositions because CS is known to show a slight variation in abundance among different types of galaxies such as starbursts, AGNs, ULIRGs, and normal galaxies \citep{mar09}. Moreover, \cite{sai22b} reported that the distribution of CS in NGC 1068 is one of the molecules that has little relation to the AGN-driven outflow. This indicates that the CS abundance is likely to be less sensitive to density variations and dissociation.

The abundant molecules in the CND with a 350 pc scale and in the AGN with a 60 pc scale relative to the SBR are in good agreement; thus, significant differences were not found based on these figures. H$^{13}$CN, SiO, HCN, and H$^{13}$CO$^{+}$ were enhanced in the entirety of all CND and AGN positions. In fact, HCN (4--3)/CS (2--1 and 7--6) in the CND and AGN position are significantly higher than that in the SBR with ALMA high-resolution ($\sim$40 pc) observations \citep{but22}. The enhancement of these molecules is possibly due to a shock by the AGN outflow. Because the line intensity ratio is increasing in high-velocity components (i.e., $>$200 km s$^{-1}$; broad wing component) than that in the central velocity components in the case of the HCN/CS intensity ratio. Unfortunately, due to low S/N, line ratios for the high-velocity components for H$^{13}$CN, SiO, and H$^{13}$CO$^{+}$ cannot be calculated. In contrast, $^{13}$CO was enhanced in the SBR with both spatial resolutions. No clear trend can be found in the other molecules because they are not much different owing to the large error bars or are not detected in the AGN and/or SBR. 

The abundance of CN, which is known as a possible tracer of X-ray dominated regions (XDRs) proposed in theoretical studies \citep[e.g.,][]{mal96, mei05, mei07}, is one of the highest among all detected molecules in the CND and AGN. Thus, strong X-ray radiation from the AGN was expected. This feature was also described in \cite{gar10}. However, CN in the SBR is also abundant based on our line survey, and the high abundance of CN is not only due to high X-ray flux, but also due to UV radiation (i.e., photon-dominated regions; PDRs) \citep{jan95}. We have already discussed the abundance of CN toward NGC 1068 in a previous work using single-dish observations \citep{nak18}, and this result does not change with the observational scale.

\begin{figure*}[h]
\plotone{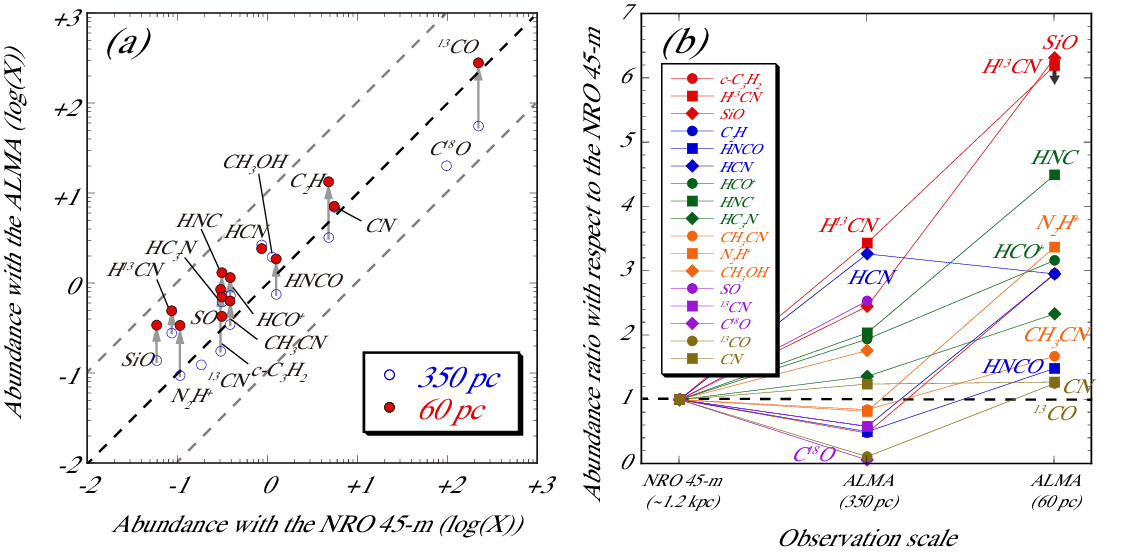}
\caption{(a) Plot of the fractional abundances with respect to CS between the observation with ALMA and the NRO 45-m telescope. Blue open circles and red filled circles represent the abundances in the CND with the 350 pc scale and those in the AGN with the 60 pc scale, respectively. These results represent the abundance correlation between the 350 pc and 1.2 kpc scales, and between the 60 pc and 1.2 kpc scales. Plots with the bold dashed line represent the same abundance without dependence on the observational scale. Arrows indicate increasing or decreasing abundance from 350 pc to the 60 pc scale. (b) Fractional abundances with respect to CS normalized by the NRO 45-m telescope traced on a 1.2 kpc scale. The bold dashed line represents an abundance ratio of unity.
\label{fig:abun_sd}}
\end{figure*}

\subsection{Comparison with the abundance using the single-dish telescope}
The NRO 45-m telescope was used to obtain molecular abundances toward the central region of NGC 1068 with an approximate 1.2 kpc scale \citep{nak18, tak19}. Figure~\ref{fig:abun_sd} shows the dependence of the scale on the molecular abundance with respect to CS by comparing ALMA results with those of the NRO 45-m telescope. It is expected that the smaller the beam size (i.e., from 1.2 kpc through 350 pc to 60pc), the clearer the effect of the AGN as an abundance enhancement or deficient. The enhancements or deficiencies are plotted above or below the lines with a slope of 1 in Figure~\ref{fig:abun_sd}(a) and lines with a ratio of 1 in (b), respectively.

In these figures, $^{13}$CO and C$^{18}$O with both 350 pc and 60 pc scales show low abundance values relative to the 1.2 kpc scale at first sight. This means the abundance with 1.2 kpc is the highest among these observational scales. The molecular gas of the CO isotopologues is distributed in the interarm region between the CND and the SBR (d $>$350 pc), as shown in \cite{tos17}, while the NRO 45-m telescope detected the emission from the components. Except for these CO isotopologues, the fractional abundance of almost all molecules at the 350 pc scale is consistent with those at the 1.2 kpc scale. The abundances of CH$_{3}$OH, H$^{13}$CN, SiO, and HCN are slightly higher (ratio $>$2), as shown in Figure~\ref{fig:abun_sd}(b). In contrast, a different trend of plots with 60 pc is clearly observed for some molecules. SiO, H$^{13}$CN, and HNC are significantly enhanced, while HCO$^{+}$, N$_{2}$H$^{+}$, C$_{2}$H, HCN, c-C$_{3}$H$_{2}$, HNCO, and HC$_{3}$N are also slightly enhanced when compared to those with the 1.2 kpc scale. Although the mechanisms of these enhancements may differ for each molecule (i.e., X-ray irradiation and/or mechanical feedback), this feature betrays an effect of the AGN on the surrounding molecular gas depending on the observational scale. The enhancement scenarios for some molecules are described in the following subsections.

\begin{figure*}
\plotone{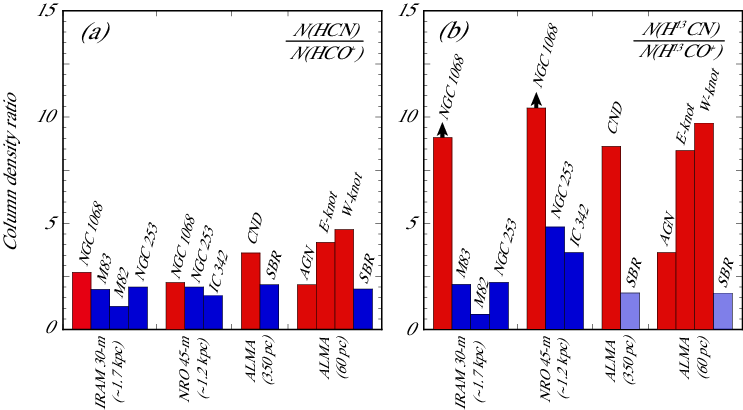}
\caption{Column density ratios of (a) $N$(HCN)/$N$(HCO$^{+}$) and (b)  $N$(H$^{13}$CN)/$N$(H$^{13}$CO$^{+}$). The values with the IRAM 30-m telescope and the NRO 45-m telescope are taken from \cite{ala15} and \cite{nak18}, respectively. Red and blue columns represent the values in the CND (or AGN) of NGC 1068 and in the SBR of NGC 1068 or starburst galaxies, respectively. Light blue columns in (b) represent the upper-limit of both the numerator and denominator of a fraction because both H$^{13}$CN and H$^{13}$CO$^{+}$ are not detected.
\label{fig:ratio}}
\end{figure*}

\subsection{Column density ratios in the AGN}
An enhancement of the HCN ($J$ = 1--0) intensity relative to HCO$^{+}$ ($J$ = 1--0) ($I$(HCN)/$I$(HCO$^{+}$)) toward the Seyfert nuclei compared with the nuclear starburst galaxies at a few 100 pc scales has been proposed in previous studies \citep[e.g.,][]{koh01, koh05}, and this trend was also confirmed in NGC 1068. \cite{koh08} reported that $R_{{\rm HCN/HCO^{+}}}$ toward the CND and the SBR in NGC 1068 are 2.1 and approximately unity, respectively, as measured with the Nobeyama Millimeter Array (NMA). Therefore, $I$(HCN)/$I$(HCO$^{+}$) may be a useful discriminator for the galactic power source between AGN and starburst activity. In addition, this trend with higher transition HCN ($J$ = 4--3) and HCO$^{+}$ ($J$ = 4--3) lines is clearer than that of the fundamental transition lines \citep{izu16}. However, the column density ratio $N$(HCN)/$N$(HCO$^{+}$) in NGC 1068 and that in typical starburst galaxies are almost the same value with single-dish telescopes \citep{ala15, nak18}. These studies reported that the values of $N$(HCN)/$N$(HCO$^{+}$) do not exhibit a clear trend corresponding to the type of galactic activity. The authors of previous studies have proposed that probable reasons for the lack of a trend are the low angular resolution and the effect of the optical depth. Moreover, \cite{nak18} proposed that the $^{13}$C isotopic species ratio (i.e., $N$(H$^{13}$CN)/$N$(H$^{13}$CO$^{+}$)) with optically thin lines, is significantly enhanced in NGC 1068 and in typical starburst galaxies with values of $>$10.4 and $\sim$4--5, respectively, and could be more useful for differentiation.

In this study, we estimate the column density ratios with the 350 and 60 pc scales, as shown in Figure~\ref{fig:ratio}. In $N$(HCN)/$N$(HCO$^{+}$) estimated at 350 pc, we determined that the ratios in the CND and SBR are 3.6 and 2.1, respectively. The value in the CND is twice as large as that in the SBR and is much different from that obtained by single-dish telescopes (Figure~\ref{fig:ratio}(a)). Therefore, a high spatial resolution is important to determine the difference in these ratios among different galaxies. In addition, we found that the ratios in the E- and W-knots are higher than those in the AGN position compared with those in the SBR with a 60 pc scale. This indicates that the ring structure surrounding the AGN at $\sim$200 pc, rather than the gas component at the AGN position, further contributes to the enhancement of $N$(HCN)/$N$(HCO$^{+}$) in the AGNs. These features are clearer in the ratio using $^{13}$C isotopic species (Figure~\ref{fig:ratio}(b)).

According to the model calculation, a high-temperature reaction ${\rm CN + H_{2} \longrightarrow HCN + H}$ efficiently converts CN into HCN, reducing the fractional abundance of CN at elevated temperatures and increasing that of HCN in hot environments \citep{har10}. Therefore, the enhancements of HCN and H$^{13}$CN in the CND may be due to the high-temperature environment. \cite{izu13} claimed that the reason for the HCN increase is mechanical heating due to the shock by the AGN jet. In fact, the E-knot is likely to be heavily shocked based on observations of the shock tracers SiO, HNCO, CH$_{3}$OH, and chemical model calculations \citep{gar10, kel17}. We also confirmed that the distribution of H$^{13}$CN in the CND is strongly correlated with the emission of SiO, which is well-known as a strong shock tracer, with additional ALMA observations (A. Taniguchi et al. in prep.). 

$N$(CN)/$N$(HCN) is expected to have a large value that lies between 40 (at $n\sim$10$^{6}$ cm$^{-3}$) and over 1000 (at $n\sim$10$^{4}$ cm$^{-3}$) in an XDR model \citep{mei07}. The density in the CND was estimated to be 10$^{5}$--10$^{6}$ cm$^{-2}$ toward the CND and AGN positions \citep{sco20}. In contrast, the estimated value in our study is approximately 15 in the CND, which is smaller (1/3--1/5) than that in the SBR. Therefore, the high-temperature environment resulting from mechanical heating in the CND is more effective than the expected strong X-ray and/or cosmic ray irradiation from the AGN based on the footprint of molecular compositions. 

However, we reported a significant enhancement of $N$(CN) in the CND compared to that in the SBR based on another ALMA observation of the higher transition line $N$ = 3--2 \citep{nak15}. In this study, higher transition lines are not used for analysis because CN emission lines are located in the band edge of the spectrometer, and the detection is only partial. For a more accurate estimation of the physical properties of these molecules, it is necessary to obtain information on multiple transitions. 

\section{Conclusions}
In this paper, we present an imaging molecular line survey in the 3-mm band (85--114 GHz) toward one of the nearest galaxies with an AGN, NGC 1068, based on observations taken with ALMA. This is the first line survey with high resolution that can resolve the internal structure of the circumnuclear disk (CND) at the nucleus in an AGN host of a nearby galaxy. The results and discussion are summarized as follows. \\

\begin{enumerate}
\item A total of 23 molecular gas distributions are obtained covering the entire structure of the central region, which consists of the CND and starburst ring (SBR), with 60 and/or 350 pc resolution. We detected a $^{29}$SiO line for the first time toward NGC 1068. Moreover, non-detection lines in the previous line survey with the NRO 45-m telescope, H$^{13}$CO$^{+}$ and CH$_{3}$CN ($J_{K}$ = 5$_{K}$--4$_{K}$), are clearly detected and imaged using the ALMA.

\item The strong emissions of H$^{13}$CN, H$^{13}$CO$^{+}$, SiO, HC$_{3}$N, CH$_{3}$CN, SO, and $^{13}$CN are concentrated in the CND, while weak emission is either seen or not seen in the SBR with 350 pc scale observations. \cite{tak14} already reported this feature for HC$_{3}$N, CH$_{3}$CN, SO, and $^{13}$CN, and we newly confirmed that other molecules have similar characteristics. The enhancement of these molecules in the CND may be a result of the effect of the AGN. In addition, CH$_{3}$OH and HNCO ($J_{Ka, Kc}$ = 5$_{0,5}$--4$_{0,4}$) have a characteristic distribution in the SBR. Their peak positions are clearly different from that of $^{12}$CO. Such a feature for CH$_{3}$OH was reported by \cite{tos17}, and we newly found that HNCO has a similar trend. These molecules may reflect dynamic effects, such as cloud-cloud collisions and/or galaxy dynamics because they are known as shock tracers.

\item The following molecules have significantly strong emissions from the E-knot relative to the W-knot: c-C$_{3}$H$_{2}$, $^{29}$SiO, H$^{13}$CO$^{+}$, CH$_{3}$CN, N$_{2}$H$^{+}$, C$^{34}$S, SO, and $^{13}$CO with 60 pc scale observations. Although there are no molecular lines of stronger emission from the W-knot than that from the E-knot, there are various peak intensity positions in the W-knot for molecules.

\item In the fractional abundances relative to CS in the CND, AGN position, and SBR, the abundant molecules in the CND with the 350 pc scale and in the AGN with the 60 pc scale are almost similar in the two resolutions. H$^{13}$CN, SiO, HCN, and H$^{13}$CO$^{+}$ show enhancements in the whole of the CND as well as the AGN position. These HCN, H$^{13}$CN, H$^{13}$CO$^{+}$, and $^{13}$CO trends were already reported using single-dish telescopes compared with the typical starburst galaxies \citep{ala13, nak18}, and ALMA observations (Butterworth et al. 2022). However, the enhancement of SiO in the CND relative to the SBR is newly found thanks to high-resolution imaging with ALMA. However, $^{13}$CO shows an enhancement in the SBR compared to that in the CND and AGN.

\item Except for the CO isotopologue, the abundance of almost all molecules with ALMA (350 pc scale) are consistent with those with the NRO 45-m telescope ($\sim$1.2 kpc). SiO and H$^{13}$CN are significantly enhanced, and HNC, N$_{2}$H$^{+}$, C$_{2}$H, HCN, c-C$_{3}$H$_{2}$, HNCO, and HC$_{3}$N are also slightly enhanced compared with those with 1.2 kpc. This may be because of the effect of AGN on surrounding molecular gas depending on the observational scale.

\item We estimate column density ratios $N$(HCN)/$N$(HCO$^{+}$) and $N$(H$^{13}$CN)/$N$(H$^{13}$CO$^{+}$) with 350 and 60 pc scales. $N$(HCN)/$N$(HCO$^{+}$) in the CND is twice as large as that in the SBR, and it is significantly different from that obtained by the single-dish telescopes. Therefore, high spatial resolution is important to determine the difference in these ratios among different galaxies. This feature is more clearly observed in the ratio using $^{13}$C isotopic species. 

\item Based on the column density ratios of $N$(HCN)/$N$(HCO$^{+}$)and $N$(CN)/$N$(HCN), the enhancements of HCN in the CND are possibly due to the high-temperature environment, which was claimed by \cite{har10} and \cite{izu13}. The expected strong X-ray and/or cosmic ray irradiation from the AGN have relatively less impact on the molecular abundance in the CND than mechanical feedback.
\end{enumerate}

\acknowledgments
This study used the following ALMA data: ADS/JAO.ALMA\#2011.0.00061.S, ADS/JAO.ALMA\\
\#2012.1.00657.S, ADS/JAO.ALMA\#2013.1.00060.S, and ADS/JAO.ALMA\#2013.1.00279.S. ALMA is a partnership of ESO (representing its member states), NSF (USA), and NINS (Japan), together with NRC (Canada), MOST and ASIAA (Taiwan), and KASI (Republic of Korea), in cooperation with the Republic of Chile. The Joint ALMA Observatory is operated by ESO, AUI/NRAO, and NAOJ. This work was supported by JSPS Grants-in-Aid for Scientific Research, KAKENHI (JP15K05031). T. S., S. T., T. N., and N. H. were supported by the NAOJ ALMA Scientific Research grant No. 2021-18A. N.H. acknowledges support from JSPS KAKENHI Grant Number JP21K03634. 

\clearpage

\appendix

\section{Rotation diagram}

\begin{figure*}[h]
\plotone{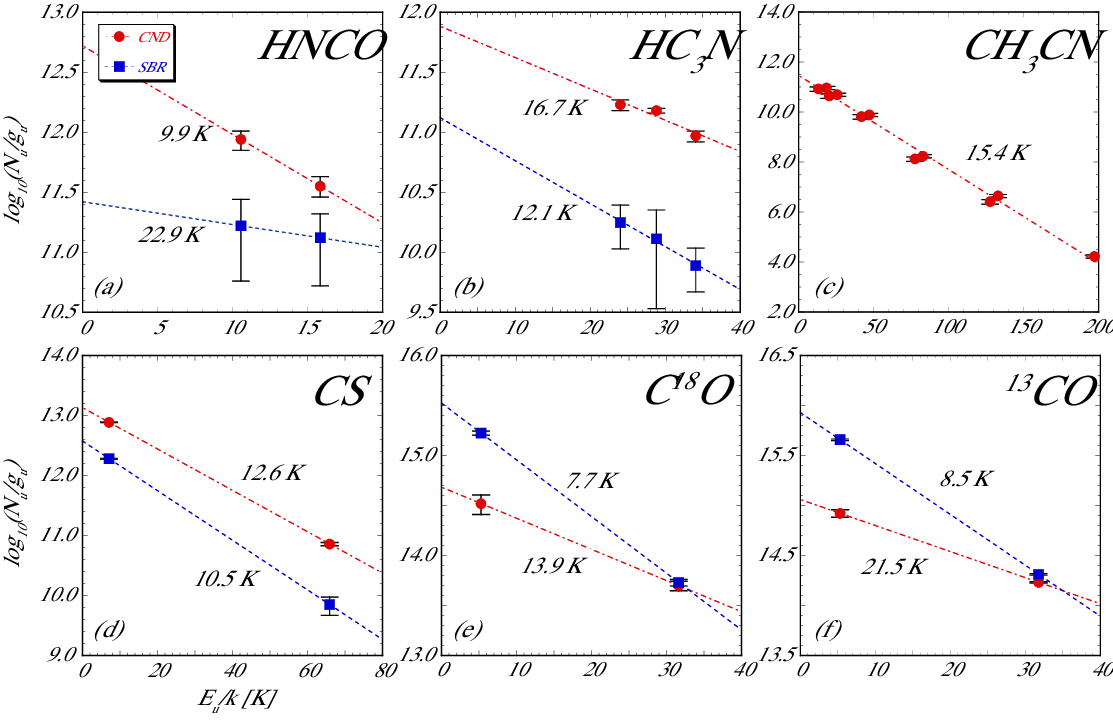}
\caption{The rotational diagram of each molecule (a) HNCO, (b) HC$_{3}$N, (c) CH$_{3}$CN, (d) CS, (e) C$^{18}$O, and (f) $^{13}$CO in the CND (circle symbols) and the SBR (square symbols). The calculated rotational temperatures are shown beside a fitting line. The error bars are based on the $\pm$1$\sigma$ noise level of integrated flux.
\label{fig:rotdia}}
\end{figure*}

The rotational temperatures ($T_{\rm rot}$) were calculated using the rotation diagram approach (Figure~\ref{fig:rotdia}) under the assumption of local thermodynamic equilibrium (LTE), that all lines are optically thin, and that a single excitation temperature characterizes all transitions (e.g., \cite{gol99}). In this study, multiple transition lines in the 3-mm band were detected in three molecules (HNCO (4$_{0,4}$--3$_{0,3}$ and 5$_{0,5}$--4$_{0,4}$); HC$_{3}$N (10--9, 11--10, and 12--11); and CH$_{3}$CN (5$_{K}$--4$_{K}$ and 6$_{K}$--5$_{K}$). For CH$_{3}$CN, we apply a method of separating blended lines based on \cite{mar06} because the transitions of different K ladders are blended. For CS, C$^{18}$O, and $^{13}$CO, we used the data in the 3-mm and 0.8-mm band \citep{nak15} to plot the rotation diagrams. Note that the errors of HNCO and HC$_{3}$N (11--10) in the SBR are too large (S/N less than 3$\sigma$); thus, we did not use the calculated $T_{\rm rot}$ of this molecule for the estimation of $N_{\rm mol}$.

\section{Column densities of molecules}

The column densities are calculated using the following equation, which is described in detail in \cite{tur91}:
\begin{equation}
{\rm log}\left(\frac{N_{\rm mol}}{Z}\right) = {\rm log}\frac{8{\pi}k{\nu}^{2}}{hc^{3}A_{ul}g_{u}g_{I}g_{K}}W + \frac{E_{u}}{k} \frac{{\rm log} \,e}{T_{\rm rot}},
\end{equation}
where $N_{\rm mol}$ is the column density, $Z$ is the partition function, $\nu$ is the frequency, $A_{ul}$ is the Einstein A-coefficient, $g_{u}$ is the rotational degeneracy of the upper state (2$J_{u}$ + 1), $g_{I}$ and $g_{K}$ are the reduced nuclear spin degeneracy and $K$-level degeneracy, respectively, and $E_{u}$ is the energy of the upper state of the transition. In this study, we obtained the value of the partition function from the Cologne Database for Molecular Spectroscopy (CDMS; \cite{mul05}). However, note that the spin degeneracy $g_{I}$ of 3 for the $^{14}$N nucleus is included in the partition functions of HCN, H$^{13}$CN, N$_{2}$H$^{+}$, CN, and $^{13}$CN. In addition, those of radicals in the 2$\Sigma$ electronic state, C$_{2}$H, CN, and $^{13}$CN, are included in the spin doublet of 2. It is necessary to consider the included degeneracies when calculating $N_{\rm mol}$ using the partition function in the database. In particular, because the degeneracies of $g_{I}$ and/or $g_{K}$ for cyclic-C$_{3}$H$_{2}$, CH$_{3}$OH and CH$_{3}$CN are complicated, we do not use the partition function in the CDMS. Details pertaining to the calculations for these molecules were explained in our previous paper \citep{nak18}.

\end{document}